\documentclass[prb,superscriptaddress,reprint,floatfix]{revtex4-2}
\usepackage{graphicx}
\usepackage{amsmath,amssymb}
\usepackage{newtxtext}
\usepackage[smallerops]{newtxmath}
\usepackage{xcolor}
\usepackage[colorlinks,bookmarks=true,citecolor=blue,linkcolor=blue,urlcolor=blue, breaklinks=true]{hyperref}
\usepackage[utf8]{inputenc}
\usepackage{t1enc}

\graphicspath{{./figures/}}

\begin{document}
\title{Noncoplanar and chiral spin states on the way towards Néel ordering in fullerene Heisenberg models}
\author{Attila Szabó} 
\address{Rudolf Peierls Centre for Theoretical Physics, University of Oxford, Oxford OX1 3PU, UK}
\address{ISIS Facility, Rutherford Appleton Laboratory, Harwell Campus, Didcot OX11 0QX, UK}
\address{Max Planck Institute for the Physics of Complex Systems, Nöthnitzer Str. 38, 01187 Dresden, Germany}
\author{Sylvain Capponi} 
\affiliation{Laboratoire de Physique Th\'eorique,
  Universit\'e de Toulouse, CNRS, UPS, 31062 Toulouse, France}
\author{Fabien Alet} 
\affiliation{Laboratoire de Physique Th\'eorique,
  Universit\'e de Toulouse, CNRS, UPS, 31062 Toulouse, France}
\date{\today}

\begin{abstract}
    Using high-accuracy variational Monte Carlo based on group-convolutional neural networks (GCNNs), we obtain the symmetry-resolved low-energy spectrum of the spin-$1/2$ Heisenberg model on several highly symmetric fullerene geometries, including the famous C$_{60}$ buckminsterfullerene.
    We argue that as the degree of frustration is lowered in large fullerenes, they display characteristic features of incipient magnetic ordering: correlation functions show high-intensity Bragg peaks consistent with Néel-like ordering, while the low-energy spectrum is organised into a tower of states.
    Competition with frustration, however, turns the simple Néel order into a noncoplanar one. 
    Remarkably, we find and predict chiral incipient ordering in a large number of fullerene structures.
\end{abstract}

\maketitle

\section{Introduction}\label{sec:intro}

Antiferromagnets on infinite bipartite lattices generally show Néel ordering in dimensions greater than one.
This can be detected through the staggered magnetisation of the ground state, which spontaneously breaks SU(2) spin-rotation symmetry.
In the spectrum, the Goldstone mode corresponding to this symmetry breaking gives rise to a proliferation of gapless states with a range of spin quantum numbers, known as the Anderson tower of states~\cite{Anderson1952TOS}, as well as a branch of gapless spin-wave excitations.
Such Néel ordering of the ground state can be proven rigorously for the Heisenberg model on the three-dimensional cubic lattice for spin $S\geq 1/2$~\cite{Dyson1978,Kennedy1988} as well as on the two-dimensional square lattice for $S\geq 1$~\cite{Neves1986}.

For two-dimensional spin-$1/2$ systems, however, and especially to study ordering in frustrated magnets, one has to rely on numerical studies that are almost always performed on finite patches of the lattice.
Spontaneous symmetry breaking never occurs in these finite systems.
Nevertheless, ordering tendencies in the thermodynamic limit are already indicated by sharp Bragg peaks (with intensity proportional to system size) in the static correlation functions,
as well as incipient Anderson towers of states in the spectrum, at energies well below those of quasiparticle excitations. 
Even beyond conventional ordering, the symmetry-resolved low-energy spectrum is an invaluable diagnostic of phases of matter, both computationally~\cite{Wietek2017TOS,wietek2017chiral,Wietek2020KagomeVBS,wietek2023triangular,Nomura2021Dirac-TypeSpectroscopy} and experimentally~\cite{Wulferding2020}. 

An interesting alternative to the paradigm above is considering strongly correlated systems on highly symmetric molecular geometries, which also exhibit a wide range of unusual quantum magnetic properties, such as magnetization jumps and plateaus, or the proliferation of lowest singlet (rather than magnetic) excitations~\cite{Schnack2010,Rousochatzakis2008,Furrer2013}.
A case in point are fullerene structures~\cite{review_fullerene}, made up of pentagonal and hexagonal faces:
while there is no limit on the number of hexagons, Euler's formula implies that they all have 12 pentagonal faces.
This allows interpolating from the limit of extreme frustration (C$_{20}$, a dodecahedron with pentagonal faces only) to large molecules that resemble the bipartite honeycomb lattice with a vanishing fraction of frustrated defects.

As a starting point to understanding strong-correlation effects in carbon fullerenes, exact-diagonalisation (ED) and quantum-Monte-Carlo studies were performed on the C$_{20}$ Hubbard model~\cite{Lin2007b}. 
These found that the spin-triplet ground state of the weakly-interacting Hückel limit switches to a nondegenerate singlet as interactions are made stronger, consistent with the Heisenberg model on the same geometry~\cite{Konstantinidis2005}.
This shows that the Heisenberg limit provides useful information about the physically more relevant~\cite{Chakravarty1991C60Hubbard,Coffey1992,Scalettar1993} intermediate-$U$ Hubbard model, which would pose considerably greater computational challenge.
The lowest-lying excitations of the C$_{20}$ Heisenberg model are also singlets, including one belonging to a five-dimensional irreducible representation (irrep) of the icosahedral point group:
the absence of low-energy triplets is incompatible with incipient magnetic ordering, as one would expect for such a highly frustrated molecule.
While ED results have been obtained for the Heisenberg model on fullerene allotropes up to C$_{36}$~\cite{Konstantinidis2009,Modine1996,Konstantinidis2018LargerED}, their high degree of frustration and varying degrees of symmetry obstructs the emergence of any systematic trend.

\begin{figure*}
    \centering
    \includegraphics{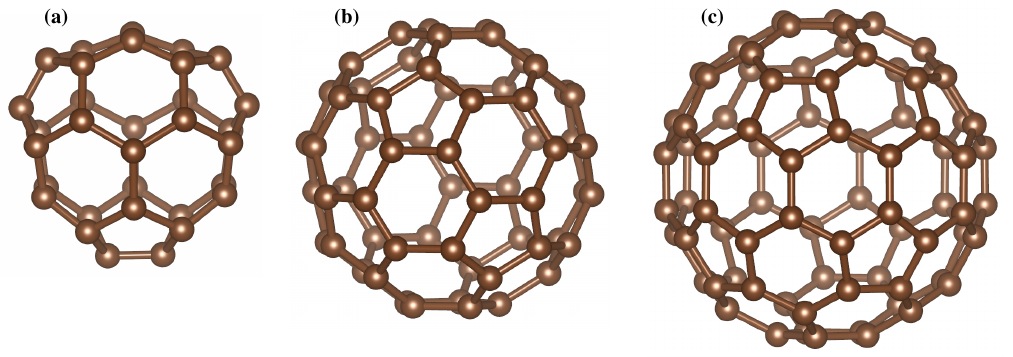}
    \caption{Connectivity of the three fullerene graphs considered in this work: (a) C$_{32}$ ($D_{3h}$ isomer), (b) C$_{60}$ ($I_h$ isomer), and (c) C$_{80}$ ($I_h$ isomer). Figures made using VESTA~\cite{Momma2011VESTAData}; coordinates of C$_{32}$ taken from Ref.~\cite{Tomanek2014NanocarbonJungle}.}
    \label{fig: molecules}
\end{figure*}

Much attention has also been devoted to the famous C$_{60}$ buckminsterfullerene geometry [Fig.~\ref{fig: molecules}(b)], motivated by interest in chemical, nanotechnology, and quantum-computing~\cite{Benjamin2006} applications, as well as superconductivity observed in alkali-metal-doped fullerene lattices~\cite{C60_superconductivity}.
Early studies of the Hubbard model~\cite{Coffey1992,Scalettar1993} argued in favour of a nondegenerate singlet ground state at strong and intermediate interaction on this geometry as well.
More recently, this has been corroborated by DMRG studies of the Heisenberg model by Rausch {\it et al.}~\cite{Rausch2021}, which found the lowest few eigenstates of the $S=0,1,2$ spin sectors.
In particular, these authors uncovered that the lowest-lying triplet excitation is threefold degenerate and breaks cubic rotation symmetry.
However, a detailed analysis of the spatial symmetries of the low-lying spectrum, desirable for understanding the low-energy physics and any symmetry-breaking tendencies of large fullerenes, requires complementary numerical approaches since, in general, tensor-network methods struggle to resolve spatial symmetries.

Here, we address this demand using a variational Monte-Carlo approach based on group-convolutional neural-network (GCNN) wave functions~\cite{Roth2021GroupAccuracy,Roth2023HighAccuracy}.
These allow us to resolve the lowest-lying states in every spatial symmetry sector with modest computational resources, and thus reconstruct much of the low-energy spectrum.
In particular, we study the spin-$1/2$ nearest-neighbour Heisenberg model
\begin{equation}
    H = J \sum_{\langle ij\rangle} \vec S_i \cdot \vec S_j
    \label{eq:Hamiltonian}
\end{equation}
on the highest-symmetry allotropes of C$_{32}$, C$_{60}$, and C$_{80}$, shown in Fig.~\ref{fig: molecules}.
In what follows, we use $J=1$ as the unit of energy and consider the zero-temperature limit.
The smallest molecule allows us to benchmark the method against ED: despite the high degree of frustration, we obtain variational energies very close to the lowest exact ones in every symmetry sector considered.
Likewise, our variational energies for C$_{60}$ match those obtained from DMRG; however, we obtain dozens of additional energies and wave functions across all icosahedral symmetry sectors.

Most importantly, we are able to account for much of this low-lying spectrum by adapting arguments on towers of states and the ground states of classical ($S\to\infty$) Heisenberg models developed for lattice models.
In particular, we find that the lowest-lying $S=0,1,2$ states are captured by a triplet of low-energy $S=1$ modes, which play the role of Goldstone modes (at gapless points of the magnon dispersion relation) of an incipient noncoplanar order.
This order can be understood as the result of the competition between incipient Néel ordering on the hexagons and frustration introduced by the pentagons:
this is highlighted by the wave functions of the Goldstone modes, which show Néel-like alternating signs on maximal bipartite segments of the C$_{60}$ geometry.
We provide a recipe, analogous to the Luttinger--Tisza method for lattice magnets, to predict the 
symmetry properties of these modes, which matches the numerical calculations perfectly.

We also perform the same analysis for the next smallest icosahedrally symmetric fullerene, C$_{80}$. 
Remarkably, its low-lying spectrum comes in nearly degenerate pairs of states, which only differ in their parity under spatial inversion.
We again account for this behaviour in terms of an incipient symmetry-breaking order.
The large-$S$ Heisenberg model on this geometry has a chiral ground state (that is, it breaks spatial inversion and time-reversal, but not their product), for which tower-of-states analysis predicts such a degeneracy.
We also construct an explicit chiral operator in terms of the Goldstone-mode operators of the incipient order to relate the pairs of states to one another.
Detecting such a chiral ordering in C$_{80}$ would be an interesting target of future computational (e.g., DMRG) and experimental studies.

The rest of the paper is organised as follows.
In Sec.~\ref{sec: ordering}, we generalise methods to detect incipient ordering in finite systems to molecules without translation symmetry.
In Sec.~\ref{sec: methods}, we describe our GCNN ansatz and its optimisation protocol in detail, and benchmark it against ED on the C$_{32}$ molecule in Sec.~\ref{sec:c32}. 
We detail our numerical studies of C$_{60}$ and C$_{80}$ in Secs.~\ref{sec:c60} and~\ref{sec:c80}, respectively.
Conjectures on the low-energy spectra of larger fullerenes, based on semiclassical arguments, are presented in Sec.~\ref{sec: large fullerene}.
Perspectives and conclusions are given in Sec.~\ref{sec: conclusion}. 
An appendix on subspace projection of high-dimensional irreps (Appendix~\ref{app: projection}) and tables of exact and variational energies (Appendix~\ref{app:data}) complete the paper. 

\section{Incipient ordering in molecular magnets}
\label{sec: ordering}

On an infinite two-dimensional lattice, the ground state of a magnetic Hamiltonian may break spin-rotation symmetry.
This is indicated by the emergence of Bragg peaks, divergences of the reciprocal-space correlation function $\langle \vec S(-\vec k_0)\cdot\vec S(\vec k_0)\rangle$ at some wave vector $\vec k_0$, as well as a gapless branch of Goldstone modes (magnons) corresponding to rotating the order parameter direction. 
The magnons become gapless at the Bragg peak position; in a primitive lattice, repeated application of the magnon creation operator
\begin{equation}
    \vec S(\vec k_0) = \sum_r e^{i\vec k_0\cdot\vec r} \vec S_r
    \label{eq: magnon lattice}
\end{equation}
to the ground state creates a sequence of zero-gap states with different spin quantum numbers, known as the \textit{(Anderson) tower of states}~\cite{Anderson1952TOS,Wietek2017TOS}.
Symmetry quantum numbers of the states in this tower can be derived from the above construction, or by decomposing symmetry-broken classical ground states into irreps of the full symmetry group $G_\mathrm{spatial}\times \mathrm{SU(2)}$~\cite{Bernu1994Triangular,Wietek2017TOS,Rousochatzakis2008}.
While no true phase transition is possible on a finite system, symmetry-breaking tendencies can readily be established from numerical simulations of finite lattices, either from the finite-size scaling of reciprocal-space correlators, or from the energy spectrum, which contains a distinct set of low-lying excitations with symmetry quantum numbers consistent with the Anderson tower of states~\cite{Wietek2017TOS}.

It is reasonable to expect similar precursors to ordering on large fullerene geometries:
Even though these are always frustrated due to having 12 pentagonal faces, in the large-molecule limit, almost every face is hexagonal, so we can regard the structure as a large honeycomb lattice with a finite number of defects. Therefore, physical properties away from these defects ought to approach those of the honeycomb lattice, which sustains Néel order~\cite{Reger89,Castro2006HoneycombQMC}.

Since the fullerene geometry has no translational symmetry, we cannot directly probe such incipient ordering in reciprocal space. 
Bragg peaks, however, can be extracted in real space as well, as the dominant eigenvector of the correlation matrix $C_{ij} = \langle \vec S_i\cdot\vec S_j\rangle$, with a diverging eigenvalue corresponding to the order parameter~\footnote{In a lattice geometry, $C_{ij}$ is translation-invariant, so these eigenvectors are plane waves, recovering the usual momentum-space treatment.}.
Likewise, the leading eigenvector of $C_{ij}$ on the fullerene geometry can be thought of as a real-space Bragg-peak ``wave function'' $\psi_i$.
This wave function can be used to construct the Goldstone-mode operator
\begin{align}
    \hat{\mathcal S}_\psi^\pm &= \sum_i \psi_i \hat S_i^\pm; &
    \hat{\mathcal S}_\psi^z &= \sum_i \psi_i \hat S_i^z, 
    \label{eq: magnon operator}
\end{align}
repeated application of which creates an ansatz ``tower of states'' that can be compared to the low-lying eigenstates of the full Hamiltonian.
Just as in the case of lattice systems, the (point-group) symmetry quantum numbers of this tower of states can be deduced either from the repeated application of the bosonic~%
\footnote[42]{Unlike an infinite lattice model, the Goldstone-mode operators~\eqref{eq: magnon operator} have a finite support on each site, so they do not perfectly commute. However, the norm of the commutator is inversely proportional to the system size, so we do not expect it to give rise to a low-lying ``antisymmetric two-Goldstone'' excitation. That is, we can treat the Goldstone modes as commuting, bosonic operators.}
operators $\hat{\mathcal{S}}_\psi$, or from decomposing a symmetry-broken classical ground state into irreps of $G_\mathrm{point}\times \mathrm{SU(2)}$, the latter of which can be made controlled in the large-$S$ limit, where such symmetry-breaking ground states may form even for finite systems~\cite{Rousochatzakis2008}.

A further analogy with lattice magnets allows us to predict these symmetry quantum numbers directly from the Hamiltonian, without computing the correlation matrix $C_{ij}$ of the quantum many-body ground state, similar to the Luttinger--Tisza method for lattice magnets~\cite{LuttingerTisza46,Lyons60,Kaplan07}.
In the large-$S$ limit underlying the above arguments, the Hamiltonian~\eqref{eq:Hamiltonian} on a lattice can be Fourier transformed,
\begin{equation}
    H = \sum_k J(k) \vec S(-k)\cdot \vec S(k),
    \label{eq: Luttinger-Tisza}
\end{equation}
without having to worry about complicated commutation relations between the $\vec S(k)$. 
The minimum of $J(k)$ predicts the position of Bragg peaks, subject to compatibility with the unit-length constraint on spins in real space, which may also determine whether the order is collinear, coplanar, or noncoplanar~\cite{Kaplan07}.
Likewise, the lowest-energy eigenvector of the Hamiltonian matrix (in our case, the adjacency matrix of the fullerene graph) is expected to recover the Bragg-peak wave function $\psi_i$.

\section{Group-convolutional neural-network states}
\label{sec: methods}

In the following section, we describe our numerical method to obtain the low-energy spectrum. 
Group-convolutional neural networks (GCNNs)~\cite{Cohen2016GroupNetworks,Roth2021GroupAccuracy,Roth2023HighAccuracy}, which play a central role in our approach, are discussed in Sec.~\ref{sec:group}; for a general discussion of neural-network-based variational Monte Carlo, we refer the reader to Refs.~\cite{Carleo2017SolvingNetworks,Carrasquilla2021}. 
In Sec.~\ref{sec: irrep projection} and Appendix~\ref{app: projection}, we explain how the symmetry of a GCNN wave function can be constrained beyond the (multi-dimensional) irreps of the point group, which we found to substantially improve our results.
Specific details of the GCNN architecture and other hyperparameters are given in Sec.~\ref{sec:details}.
Finally, benchmarks against ED on a C$_{32}$ allotrope are presented in Sec.~\ref{sec:c32}.

\subsection{Ansatz}
\label{sec:group}

Space-group symmetries (that is, ones that map computational basis states onto one another) can be imposed on any variational ansatz $\psi_0$ using the projection formula~\cite{Heine1960GroupMechanics}
\begin{subequations}
\begin{align}
    |\psi\rangle &= \frac{d_\chi}{|G|}\sum_{g\in G} \chi_g^* \hat g|\psi_0\rangle\\
    \psi(\sigma)\equiv \langle \sigma|\psi\rangle &= \frac{d_\chi}{|G|}\sum_{g\in G} \chi_g^* \psi_0\left[\hat g^{-1}(\sigma)\right],
    \label{eq: irrep projection basis states}
\end{align}
\label{eq: irrep projection formula}%
\end{subequations}
where the $\hat g$ are the elements of the space group $G$ and the $\chi_g$ are their characters in a given $d_\chi$-dimensional irrep of $G$. Here $\sigma$ stands for a spin configurationin the computational $S^z$ basis: $\sigma=( S^z_1, S^z_2 \dots S^z_N )$ for $N$ spins-$1/2$ with $S^z_i \in \{\uparrow, \downarrow \}\equiv \{\pm1\}$.
Applying (\ref{eq: irrep projection basis states}) directly, however, requires evaluating the ansatz $\psi_0$ many times, which may be prohibitively computationally expensive.

Instead, we use \textit{group-convolutional neural networks (GCNNs)}~\cite{Cohen2016GroupNetworks,Roth2021GroupAccuracy,Roth2023HighAccuracy},
a generalisation of the well-known convolutional neural networks (CNNs) to non-translational symmetries, which are able to efficiently generate all symmetry-related evaluations of a neural-network ansatz as hidden layers indexed by the symmetry elements.
Our feed-forward GCNNs start with an embedding layer
\begin{subequations}
\begin{equation}
    h^{(1)}_g = \sum_{\vec r} K(\hat g^{-1} \vec r) \sigma(\vec r),
    \label{eq: embedding layer}
\end{equation}
which converts the input spin configuration $\sigma$ into such a hidden layer, $h^{(1)}$. Then, further group-valued hidden layers are generated by alternating nonlinearities and equivariant linear layers of the form
\begin{equation}
    h^{(i+1)}_g = \sum_{k\in G} W^{(i)} (k^{-1}g) h^{(i)}_k
    \label{eq: GC layer}.
\end{equation}
The trainable variables of the ansatz are the kernel entries $K(\vec r)$ and $W(g)$.
\end{subequations}
One can show~\cite{Vicentini2022NetKetSystems} that acting with a symmetry element $\hat k$ on the input spin configuration $\sigma$ permutes the labels of all subsequent layers as
\begin{equation}
    h_g\left[\hat k^{-1}(\sigma)\right] = h_{kg}(\sigma).
\end{equation}
Therefore, we can regard the entries of the last layer as the amplitudes of a neural-quantum-state (NQS) ansatz $h^{(L)}_0(\sigma)$ evaluated for all spin configurations related to $\sigma$ by space-group symmetry:
\begin{equation}
    h^{(L)}_g(\sigma) = h^{(L)}_0\left[\hat g^{-1}(\sigma)\right],
\end{equation}
so a symmetric ansatz is obtained by combining all entries of the last layer according to the projection formula~\eqref{eq: irrep projection formula}.

In addition to using spatial symmetries of the molecules and the conservation of the magnetization  $S^z=\sum_{i=1}^N S_i^z$, the parity symmetry $\hat P = \prod_{i=1}^N S_i^x$, an element of the SU(2) spin-rotation group, can be implemented in the $S^z$ computational basis by flipping the sign of all $S_i^z$. 
We therefore imposed eigenvalues of $P=\pm1$ on our ansätze in addition to the space-group irreps.
Sampling in the $S^z=0$ magnetisation sector, this allows us to distinguish between states with even ($P=+1$) and odd ($P=-1$) total-spin quantum number.
We also performed simulations in the $S^z=2$ sector, which isolate total-spin quantum numbers $S\ge2$.

\begin{figure}
    \centering
    \includegraphics{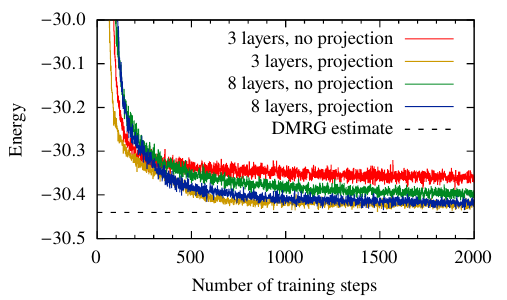}
    \caption{Evolution of the variational energy during the training protocol for three- and eight-layer GCNN ansätze, both with and without the irrep subspace projection described in Sec.~\ref{sec: irrep projection}, for the $\mathrm{H_g}$ ($P=+1$) irrep of the C$_{60}$ fullerene structure. The energies are compared to the second-lowest spin-singlet energy found in Ref.~\cite{Rausch2021} using DMRG (cf.~Fig.~\ref{fig:c60-summary}).}
    \label{fig: Hg+}
\end{figure}

\subsection{Irrep subspace projection}
\label{sec: irrep projection}

In our numerical experiments, we found that training ansätze projected on higher-dimensional ($d_\chi>1$) irreps directly using~\eqref{eq: irrep projection formula} is slower, less reliable, and more liable to instabilities than one-dimensional irreps, as shown in Fig.~\ref{fig: Hg+}.
This may be caused either by the training ``wandering'' between different wave functions in symmetry-protected multiplets, or by the zero characters $\chi_g=0$ typical in these irreps reducing the number of wave-function terms in the sum~\eqref{eq: irrep projection formula}, which is known to limit the expressivity of NQS ansätze~\cite{Reh2023NQSdesign}.
We remedied this problem by imposing further symmetry constraints that select a unique representative of each symmetry multiplet.
Effectively, we project our wave functions first on the trivial irrep of a subgroup of $G$, followed by projecting on the desired irrep of $G$ itself.
As explained in Appendix~\ref{app: projection}, the combined effect of these projections can still be written in the form~\eqref{eq: irrep projection formula} with an effective character $\tilde\chi$, which is no longer an irrep character of $G$, but has overlap with precisely one of them.
The benefits of this approach are illustrated for the five-dimensional $\mathrm{H_g}$ ($P=+1$) irrep of C$_{60}$ in Fig.~\ref{fig: Hg+}, which shows that subspace projection allows the variational optimiser to reach lower energies in fewer iterations.

\subsection{Details of the numerical experiments}
\label{sec:details}

\begin{figure}
    \centering
    \includegraphics{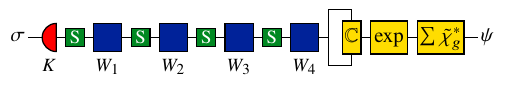}
    \caption{Structure of a five-layer GCNN of the type used in this work. Red and blue boxes stand for the embedding layer~\eqref{eq: embedding layer} and the group convolutions~\eqref{eq: GC layer}, respectively. Green boxes indicate \textsc{selu} activation functions. Yellow boxes represent the output layer~\eqref{eq: detailed ansatz}. Figure based on Ref.~\cite{Roth2023HighAccuracy}.}
    \label{fig: ansatz}
\end{figure}

To obtain the results reported below,  we used the same GCNN architecture as Ref.~\cite{Roth2023HighAccuracy}, illustrated in Fig.~\ref{fig: ansatz}. 
We use real-valued kernels $K,W$, interspersed with \textsc{selu} activation functions, which allow us to reliably train deep GCNNs~\cite{selu2017,Roth2023HighAccuracy}.
In the output layer, we combine pairs of feature maps into complex-valued features, exponentiate them, and project the result on the desired irrep:
\begin{subequations}
\begin{align}
    \tilde h_{n,g} &= h^{(L)}_{n,g} + ih^{(L)}_{n+F/2,g} \\
    \psi(\sigma) &= \sum_{g\in G} \chi_g^* \sum_{n=1}^{F/2} \exp(\tilde h_{n,g}),
\end{align}
\label{eq: detailed ansatz}%
\end{subequations}
where $L$ is the number of hidden layers and $F$ is the number of (real-valued) feature maps.
Including exponentiation in~\eqref{eq: detailed ansatz} is important to represent the wide dynamical range of wave-function amplitudes.

We used GCNNs with eight hidden layers, each composed of 32 (for the C$_{32}$ geometry) or 12 (for the C$_{60}$, C$_{80}$ geometries) feature maps,
containing 174\,336, 243\,456, and 243\,936 real variational parameters for the C$_{32}$, C$_{60}$, and C$_{80}$ geometries, respectively.
We also compare the performance of these networks with shallower (three-layer) ones in Fig.~\ref{fig: Hg+} for the second-lowest-energy spin-singlet state of the C$_{60}$ Heisenberg model, which transforms under the $\mathrm{H_g}$ irrep of the $I_h$ point group (cf.~Sec.~\ref{sec:c60}). 
Without the irrep subspace projection described in Sec.~\ref{sec: irrep projection}, increasing network depth leads to a significant improvement in variational energy; however, after applying the projection, both GCNNs equally outperform the unprojected eight-layer one.

The ansätze were trained on a single A100 GPU using the stochastic reconfiguration algorithm implemented in NetKet~\cite{Vicentini2022NetKetSystems} with learning rate $\eta=0.02$.
To maximally exploit GPU parallelism, we used 1024 parallel Markov chains to generate 3072 Monte Carlo samples per training step.
In most simulations, we performed 2000 training steps, which took between 7 (for C$_{32}$) and 28 (for C$_{80}$) GPU hours.

The variational energies and spin correlation functions reported below were obtained from averaging VMC local estimators of the Hamiltonian and the operators $\vec{S}_i \cdot\vec{S}_j$ for every pair of sites $i,j$, obtained for the same set of $2^{18}=262\,144$ samples for all operators.
For wave functions projected on one-dimensional irreps, we expect that spin correlators across symmetry-related pairs of spins are equal:
therefore, we explicitly averaged these correlators for the plots below.
In addition, we summed the local estimators of $\vec{S}_i\cdot\vec{S}_j$ for all pairs of sites to obtain an estimate of the total-spin expectation value~$\langle {S}^2\rangle$.

\subsection{C\textsubscript{32}: comparison to exact diagonalisation}
\label{sec:c32}

\begin{figure}
    \includegraphics{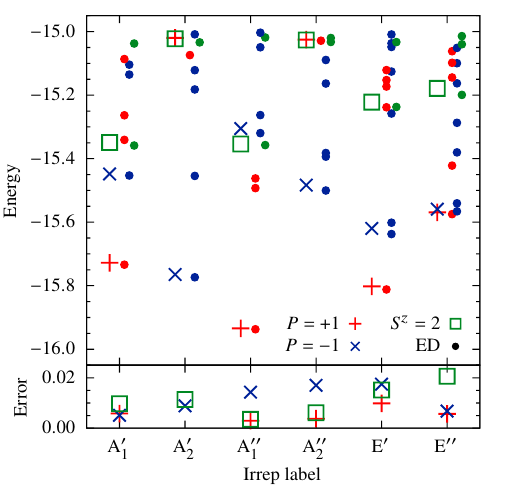}
    \caption{Best GCNN variational energies for the C$_{32}$ geometry (symbols), compared with the low-energy spectrum obtained from exact diagonalisation (red, blue, and green dots for $S=0,1,2$, respectively). The exact and variational energies are also given in Tables~\ref{tab: C32 ED} and~\ref{tab: C32 energy} of Appendix~\ref{app:data}.
    Lower panel: difference between the variational and lowest exact energies in each sector. }
    \label{fig:c32-summary}
\end{figure}

We first benchmark our method on the highest-symmetry ($D_{3h}$) isomer of C$_{32}$, labelled II in Ref.~\cite{Konstantinidis2009}, where exact diagonalisation is still possible. 
We extended the exact spectrum in Ref.~\cite{Konstantinidis2009} to all eigenstates below energy $-15$: 
The energies and point-group and spin quantum numbers of these states are listed in Table~\ref{tab: C32 ED} in Appendix~\ref{app:data}.
The best variational energies achieved using the GCNN ansatz are listed in Table~\ref{tab: C32 energy} and plotted against the exact spectrum in Fig.~\ref{fig:c32-summary}.
In every point-group and parity symmetry sector, we achieve excellent agreement with the exact results (see bottom panel with difference between exact and variational energies), with variational energies approaching the exact ground states much closer than the first excited state in the given symmetry sector.

\begin{table}[]
    \centering
    \setlength{\tabcolsep}{0.75em}
    \begin{tabular}{clll} \hline\hline
        Irrep & \multicolumn{1}{c}{$P=+1$} & \multicolumn{1}{c}{$P=-1$} & \multicolumn{1}{c}{$S^z=2$} \\\hline
        $\mathrm{A_1'}$	&	$0.0075(6)$	&	$2.0054(5)$	&	$6.0065(5)$	\\
        $\mathrm{A_2'}$	&	$5.9338(15)$\textsuperscript{a}	&	$2.0072(7)$	&	$6.0095(7)$	\\
        $\mathrm{A_1''}$	&	$0.0036(4)$	&	$2.0131(9)$	&	$6.0032(4)$	\\
        $\mathrm{A_2''}$	&	$5.9975(7)$\textsuperscript{a}	&	$2.0187(10)$	&	$6.0060(5)$	\\
        $\mathrm{E'}$	&	$0.0116(7)$	&	$2.0066(5)$	&	$6.0125(8)$	\\
        $\mathrm{E''}$	&	$0.0085(6)$	&	$2.0054(6)$	&	$6.0184(9)$	\\ \hline\hline
        \multicolumn{4}{l}{\footnotesize\textsuperscript{a}$P=+1$ simulation that returned an $S=2$ state.}
    \end{tabular}
    \caption{Total $\langle {S}^2\rangle$ for the optimised GCNN wave functions on the C$_{32}$ geometry. All are close to $S(S+1)$ for an integer spin quantum number $S$, indicating an accurately spin-rotation-symmetric wave function. 
    }
    \label{tab: C32 S2}
\end{table}

Estimates of the total spin $\langle S^2\rangle$ for our optimised wave functions are listed in Table~\ref{tab: C32 S2}. 
In every symmetry sector, we obtain a value extremely close to $S(S+1)$ for an integer spin quantum number $S$, as expected for a fully spin-rotation-symmetric state.
Every odd-parity and $S^z=2$ simulation returned states consistent with $S=1$ and $S=2$, respectively;
in the even-parity sector, we find $S=2$ ground states in two point-group symmetry sectors.
In these sectors, we also find that the optimal variational energies in the $P=+1$ and $S^z=2$ sectors coincide to very good accuracy, indicating that the two wave functions are the $S^z=0$ and $S^z=2$ states of the same quintet.

\begin{figure}
    \centering
    \includegraphics{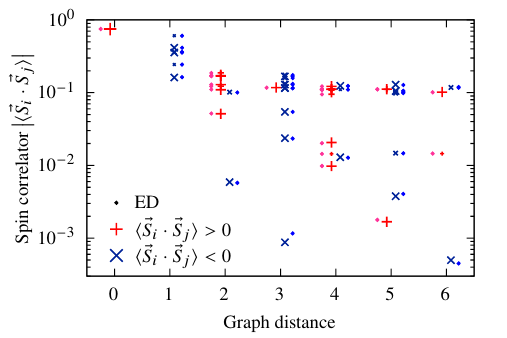}
    \caption{Ground-state spin--spin correlation functions $\langle \vec{S}_i\cdot\vec{S}_j\rangle$ as a function of graph distance on the C$_{32}$ geometry from the GCNN simulation (symbols) and ED (dots). The size of the symbols is proportional to the number of symmetry-related paths with equal correlators.}
    \label{fig: C32 corr}
\end{figure}

Finally, spin correlation functions $\langle \vec{S}_i\cdot\vec{S}_j\rangle$ for the GCNN estimate of ground state are shown in Fig.~\ref{fig: C32 corr} compared to the correlators of the exact ground state (cf.~Ref.~\cite{Konstantinidis2009}); the two again match excellently.
Due to the relatively low symmetry of the C$_{32}$ geometry, there are many inequivalent lattice sites and paths at all graph distances, so we limit ourselves to plotting the correlators as a function of graph distance only.
We find no clear pattern in the correlators:
their signs deviate from N\'eel order already for next-nearest neighbours,
and their magnitudes are spread over a wide range of values at any given graph distance.
This is consistent with the strong frustration expected in this molecule with 12 pentagonal and only 6 hexagonal faces.

\section{Incipient noncoplanar order in C\textsubscript{60}}
\label{sec:c60}

\begin{figure}
    \includegraphics{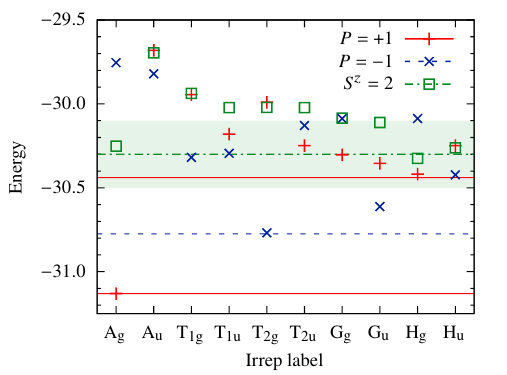}
    \caption{Best GCNN variational energies for the C$_{60}$ geometry (symbols have the same meaning as in Fig.~\ref{fig:c32-summary}) compared to optimal energies in SU(2)-symmetric DMRG~\cite{Rausch2021} (horizontal lines). The error bars on the $S=2$ DMRG result are represented by the green background shading.}
    \label{fig:c60-summary}
\end{figure}

\begin{table}[]
    \centering
    \setlength{\tabcolsep}{0.75em}
    \begin{tabular}{clll} \hline\hline
        Irrep & \multicolumn{1}{c}{$P=+1$} & \multicolumn{1}{c}{$P=-1$} & \multicolumn{1}{c}{$S^z=2$} \\\hline
        $\mathrm{A_g}$	&	$0.0022(3)$	&	$2.0121(8)$	&	$6.0038(4)$	\\
        $\mathrm{A_u}$	&	$\mathit{5.530(3)}$\textsuperscript{a}	&	$2.0541(15)$	&	$6.0279(10)$	\\
        $\mathrm{T_{1g}}$	&	$6.0013(9)$\textsuperscript{a}	&	$2.0078(6)$	&	$6.0542(14)$	\\
        $\mathrm{T_{1u}}$	&	$0.0119(7)$	&	$2.0067(6)$	&	$6.0283(11)$	\\
        $\mathrm{T_{2g}}$	&	$5.9898(16)$\textsuperscript{a}	&	$2.0065(7)$	&	$6.0129(8)$	\\
        $\mathrm{T_{2u}}$	&	$0.0105(7)$	&	$2.0347(12)$	&	$6.0295(12)$	\\
        $\mathrm{G_g}$	&	$0.0241(10)$	&	$2.0356(14)$	&	$6.0105(7)$	\\
        $\mathrm{G_u}$	&	$0.0230(10)$	&	$2.0104(7)$	&	$6.0088(6)$	\\
        $\mathrm{H_g}$	&	$0.0524(14)$	&	$2.0480(15)$	&	$6.0101(7)$	\\
        $\mathrm{H_u}$	&	$5.9873(13)$\textsuperscript{a}	&	$2.0095(7)$	&	$6.0166(8)$	\\
        \hline\hline
        \multicolumn{4}{l}{\footnotesize\textsuperscript{a}$P=+1$ simulation that returned an $S=2$ state.}
    \end{tabular}
    \caption{Total $\langle {S}^2\rangle$ for the optimised GCNN wave functions on the C$_{60}$ geometry. All but one (in italics) are very close to $S(S+1)$ for an integer spin quantum number $S$, indicating an accurate spin-rotation-symmetric wave function.}
    \label{tab: C60 S2}
\end{table}

Next, we consider the highest ($I_h$) symmetry isomer of C$_{60}$, the famous buckminsterfullerene geometry.
The optimised VMC energies for both parities, as well as in the $S^z=2$ sector, are shown in Fig.~\ref{fig:c60-summary} (see also Table~\ref{tab: C60 energy} in the appendix), while the expectation values of $S^2$ are listed in Table~\ref{tab: C60 S2}.
Similar to the C$_{32}$ geometry, they are consistent with fully spin-rotation-symmetric states.
In four symmetry sectors, the lowest-lying even-parity state is not a singlet but a quintet:
This is also evidenced by the near coincidence of the optimised energies for $P=+1$ and $S^z=2$.

Our variational energies match those of the lowest-energy $S=0,1,2$ states, as well as the first excited $S=0$ state, found in a recent SU(2)-symmetric DMRG study~\cite{Rausch2021}.
The ground state is found to transform under the trivial irrep of the $I_h$ point group.
In agreement with DMRG, the first excited state is a spin-triplet;
the $\mathrm{T_{2g}}$ irrep we identify is also qualitatively consistent with the cubic-symmetry-breaking pattern seen in the DMRG wave function.
The lowest-energy $S=2$ state is in the $\mathrm{H_g}$ irrep, with $\mathrm{A_g}$ and $\mathrm{H_u}$ states at only slightly slightly higher energies~\footnote{This coincidence of 11 states at nearby energies may explain the large error bars of the corresponding DMRG result, together with the smaller bond dimension (compared to the $S^z=0$ states) used in the calculation of Ref.~\cite{Rausch2021}}.

\begin{figure}
    \centering
    \includegraphics{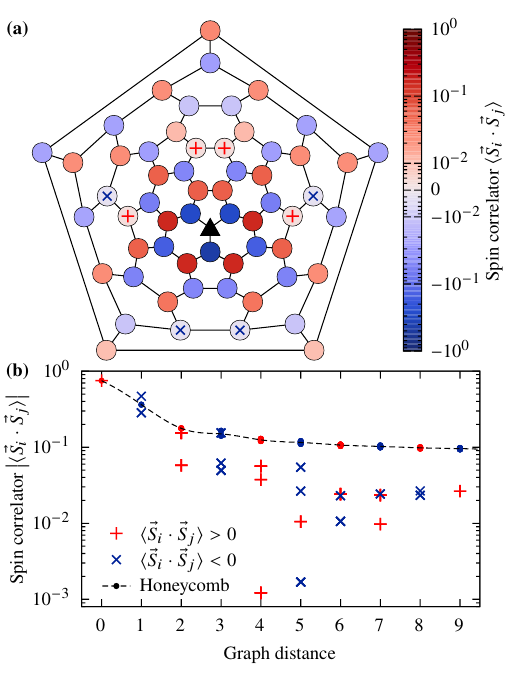}
    \caption{(a) Ground-state spin--spin correlation functions $\langle \vec{S}_i\cdot\vec{S}_j\rangle$ in the C$_{60}$ geometry. The reference point $i$ is marked with a black triangle. Two values below 0.005 in magnitude (highlighted with coloured symbols) were truncated for visibility.
    (b) Spin--spin correlators as a function of graph distance. Red and blue symbols stand for positive and negative correlators, respectively. Coloured dots show the spin correlation functions of a 512-site honeycomb lattice, measured using QMC; the dashed line is a spline connecting these dots and is included as a guide to the eye.}
    \label{fig: C60 correlation}
\end{figure}

The ground-state spin correlation function $\langle \vec{S}_i \cdot\vec{S}_j\rangle$ is shown in Fig.~\ref{fig: C60 correlation}.
Unlike C$_{32}$, every site of the buckminsterfullerene geometry is equivalent, allowing us to also display the spatial distribution of the correlators.
Our results are very close to the correlation functions measured in DMRG~\cite{Rausch2021}.
For short graph distances, they follow an alternating sign pattern consistent with Néel ordering, and their amplitudes are close to those on the unfrustrated honeycomb lattice at the same graph distances, which we computed using stochastic series expansion~\cite{Sandvik1991} with the ALPS library~\cite{AletSSE,Alps2}. 
Further away, frustration reduces correlators and introduces a nontrivial sign structure, which, somewhat surprisingly, matches that of the classical ground state discussed in Ref.~\cite{Coffey1992} (cf.\ their Fig.~1).
[For eight sites, marked with symbols in Fig.~\ref{fig: C60 correlation}(a), the classical correlator is zero, while the $S=1/2$ correlator is anomalously low.]
Remarkably, at the largest graph distances, we again recover a N\'eel-like pattern with amplitudes around $\pm0.02$; however, their signs are inverted compared to the honeycomb lattice.

\begin{figure}
    \centering
    \includegraphics{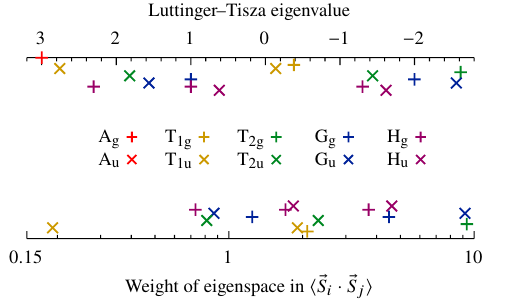}
    \caption{Top: eigenvalues of the classical Hamiltonian matrix (i.e., the adjacency matrix) in the C$_{60}$ geometry. Note that the axis is reversed, so the lowest eigenvalue (the Luttinger--Tisza ground state) is to the right. Bottom: log-scale plot of the weight (eigenvalue times degeneracy) of the eigenspaces of the spin correlator matrix $\langle \vec S_i \cdot \vec S_j\rangle$. The eigenvalue of one $\mathrm{A_g}$ eigenvector, corresponding to the net magnetisation, is zero within Monte Carlo error and is not plotted.}
    \label{fig: C60 LT spectrum}
\end{figure}

To extract signatures of ordering from our data, we computed the eigenvalues of the correlation matrix $C_{ij} = \langle \vec S_i\cdot\vec S_j\rangle$, 
as well as the the adjacency matrix of the C$_{60}$ graph, which plays the role of the Hamiltonian matrix in the Luttinger--Tisza method.
These spectra are plotted in Fig.~\ref{fig: C60 LT spectrum}.
The ground state of the Luttinger--Tisza Hamiltonian is threefold degenerate, forming a $\mathrm{T_{2g}}$ irrep of $I_h$.
This is compatible with the unit-length constraint of classical spins: the $S^x,S^y,S^z$ components of the ground state form an orthonormal basis of the irrep, leading to a \emph{noncoplanar} ground state, which can indeed be found by numerically minimising the classical Hamiltonian~\cite{Coffey1992}.
Due to quantum fluctuations, the spectrum of the spin-$1/2$ correlation matrix $C_{ij}$ does not only contain this irrep;
however, the weight (that is, the eigenvalue times the irrep dimension) of other irreps is suppressed roughly exponentially in the classical energy cost (note the logarithmic scale in the bottom panel of Fig.~\ref{fig: C60 LT spectrum}). 
In particular, $C_{ij}$ is dominated by the two lowest-energy irreps of the classical Hamiltonian, $\mathrm{T_\mathrm{2g}}$ and $\mathrm{G_u}$.
The two lowest-lying spin-triplet states also belong to these irreps, as expected from the tower-of-states construction of Sec.~\ref{sec: ordering}.
In particular, we find that the overlap between the state $\hat{\mathcal{S}}^z|\mathrm{GS}\rangle$, generated by the Goldstone-mode operator~\eqref{eq: magnon operator} applied to the ground-state $|\mathrm{GS}\rangle$, and the lowest-energy $\mathrm{T_{2g}}$ triplet ($S^z=0$, odd parity) state is $\approx 0.917$, very high for two 60-spin many-body states.

Applying the tower-of-states analysis introduced in Refs.~\cite{Wietek2017TOS,Rousochatzakis2008} to the noncoplanar classical ground state correctly predicts that the lowest-lying $S=2$ state transforms under the $\mathrm{H_g}$ irrep. 
We can also reach this conclusion by applying the bosonic~\cite{Note42} Goldstone-mode operators~\eqref{eq: magnon operator} twice to the ground state.
There is a total of nine such operators (threefold spatial and spin degeneracy), so the two-Goldstone Hilbert space consists of $9\times10/2=45$ states, transforming under the symmetric square of the $\mathrm{T_{2g}} \otimes (S=1)$ irrep of $I_h\times\mathrm{SU(2)}$:
\begin{align}
    \mathrm{Sym}^2[\mathrm{T_{2g}} \otimes (S=1)] &= (\mathrm{A_g}\oplus \mathrm{H_g})\otimes [(S=0)\oplus (S=2)] \nonumber\\
    &\oplus \mathrm{T_{2g}} \otimes (S=1).
\end{align}
The $\mathrm{A_g}$ singlet and the $\mathrm{T_{2g}}$ triplet cannot be distinguished from the ground state and the one-Goldstone state based on symmetry quantum numbers; in fact, we expect them to have a high overlap. 
The $\mathrm{H_g}$ singlet and the $\mathrm{A_g}$ quintet, however, appear in the spectrum nearly degenerate with the $\mathrm{H_g}$ quintet, even though they are not predicted by the tower-of-states analysis.
The similarly low-energy $\mathrm{H_u}$ quintet cannot be explained based on $\mathrm{T_{2g}}$ operators alone, but it may come from a combination of low-lying $\mathrm{T_{2g}}$ and $\mathrm{G_u}$ triplet excitations.

\begin{figure}
    \centering
    \includegraphics{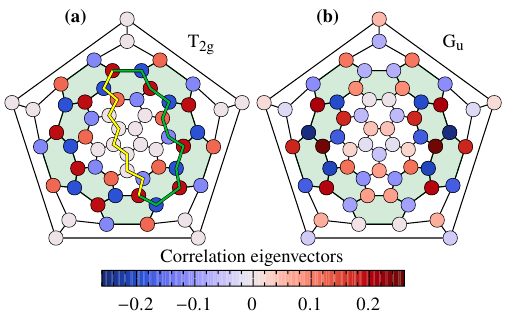}
    \caption{Eigenvectors of the ground-state correlator matrix $\langle \vec{S}_i \cdot\vec{S}_j \rangle$ in the C$_{60}$ geometry corresponding to its largest [(a); $\mathrm{T_{2g}}$ irrep; 20.7\% of all correlations] and second largest [(b); $\mathrm{G_u}$ irrep; 20.4\%] eigenvalues.
    The ten hexagons on the ``equator'' of the C$_{60}$ structure are highlighted in green. The yellow line in panel (a) indicates the shortest path (9 steps) connecting two antipodal points; the green line is the shortest path (10 steps) passing through sites with nonzero amplitude in the eigenvector.
}
    \label{fig: C60 PCA}
\end{figure}

A representative (maximally symmetric around the centre of the Schlegel plot) eigenvector of the dominant $\mathrm{T_{2g}}$ irrep of $C_{ij}$ is plotted in Fig.~\ref{fig: C60 PCA}(a). 
This eigenvector follows a perfect Néel pattern on ten hexagons around the ``equator'' of the C$_{60}$ structure, which is in fact its largest unfrustrated portion:
this indicates a clear tendency towards the Néel ordering expected in the limit of large molecules.
Away from these hexagons, frustration causes the eigenvector components to vanish.
The concentration of the Goldstone wave function around the equator of the molecule matches the distribution of ``local spin'' in the lowest-energy triplet state obtained in DMRG~\cite{Rausch2021}. 
Eigenvectors in the next-highest-weight $\mathrm{G_u}$ irrep [Fig.~\ref{fig: C60 PCA}(b)] live mostly on the same set of hexagons and display a Néel pattern modulated with a standing wave, analogous to a long-wave magnon excitation on a lattice model.

The structure of the leading correlation eigenvector explains some surprising features of the spin correlators in Fig.~\ref{fig: C60 correlation}, in particular the sign inversion of the ``Néel order'' seen at the largest graph distances.
These correspond to pairs of sites belonging to opposite hexagons:
Correlations between them are mostly mediated through the ``equatorial belt'' where the leading correlation eigenvectors are located.
Along this belt, antipodal sites are 10 steps apart [see, e.g., the green path in Fig.~\ref{fig: C60 PCA}(a)], therefore, they have the same sign in the Néel pattern of the eigenvector, consistent with their positive correlation shown in Fig.~\ref{fig: C60 correlation}.
However, the shortest path [e.g., the yellow path in Fig.~\ref{fig: C60 PCA}(a)] between the same points only takes nine steps, so we observe positive correlations at odd graph distance, in an apparent inversion of the Néel pattern.
However, since all of these odd-length paths pass through the fully-frustrated central, low-weight, region of the Schlegel plot, they do not contribute to the correlation function.
Likewise, neighbours of the antipodal point are nine (eight) steps apart along the equatorial belt (frustrated region), which extends the inverted Néel pattern to this graph distance too.

\section{Chiral inversion-symmetry breaking in C\textsubscript{80}}
\label{sec:c80}

\begin{figure}
    \includegraphics{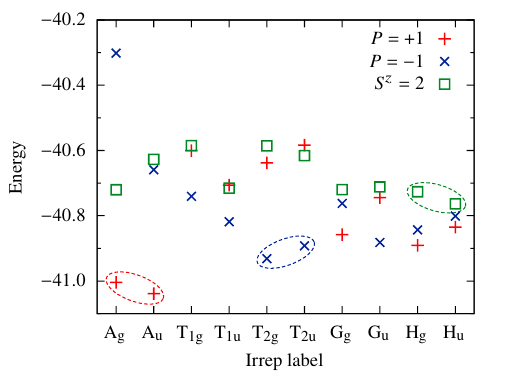}
    \caption{Best GCNN variational energies for the C$_{80}$ geometry.
    Symbols have the same meaning as in Figs.~\ref{fig:c32-summary} and \ref{fig:c60-summary}.
    The two lowest-energy states in each spin-quantum-number sector are circled with dashed lines.
    }
    \label{fig:c80-summary}
\end{figure}

\begin{table}[]
    \setlength{\tabcolsep}{0.75em}
    \begin{tabular}{clll} \hline\hline
        Irrep & \multicolumn{1}{c}{$P=+1$} & \multicolumn{1}{c}{$P=-1$} & \multicolumn{1}{c}{$S^z=2$} \\\hline
        $\mathrm{A_g}$	& $0.0185(9)$	& $2.242(3)$	& $6.0209(10)$ \\
        $\mathrm{A_u}$	& $0.0119(7)$	& $2.0248(11)$	& $6.0264(11)$ \\
        $\mathrm{T_{1g}}$	& $6.019(2)$\textsuperscript{a}	& $2.233(3)$	& \!\!\!$11.766(3)$\textsuperscript{b} \\
        $\mathrm{T_{1u}}$	& $5.958(2)$\textsuperscript{a}	& $2.153(3)$	& $6.064(2)$ \\
        $\mathrm{T_{2g}}$	& $1.998(6)$\textsuperscript{c}	& $2.0301(13)$	& $6.093(2)$ \\
        $\mathrm{T_{2u}}$	& $6.010(2)$\textsuperscript{a}	& $2.075(2)$	& $6.164(2)$ \\
        $\mathrm{G_g}$	& $0.466(3)$	& $2.118(2)$	& $6.085(2)$ \\
        $\mathrm{G_u}$	& $0.704(4)$	& $2.179(3)$	& $6.122(2)$ \\
        $\mathrm{H_g}$	& $0.343(3)$	& $2.0398(14)$	& $6.147(2)$ \\
        $\mathrm{H_u}$	& $0.350(3)$	& $2.093(2)$	& $6.091(2)$ \\\hline\hline
        \multicolumn{4}{l}{\footnotesize\textsuperscript{a}$P=+1$ simulation that returned an $S=2$ state.}\\[-1ex]
        \multicolumn{4}{l}{\footnotesize\textsuperscript{b}$S^z=2$ simulation that returned an $S=3$ state.}\\[-1ex]
        \multicolumn{4}{l}{\footnotesize\textsuperscript{c}State not clearly dominated by one $S$-sector.}
    \end{tabular}
    \caption{Total $\langle {S}^2\rangle$ for the optimised GCNN wave functions on the C$_{80}$ geometry.}
    \label{tab: C80 S2}
\end{table}

After buckminsterfullerene, the smallest fullerene structure with full icosahedral symmetry is the 80-site molecule shown in Fig.~\ref{fig: molecules}(c). 
The converged variational energies in all symmetry sectors are shown in Fig.~\ref{fig:c80-summary} and Table~\ref{tab: C80 energy}; 
interestingly, the ground-state is found in a nontrivial point-group irrep (namely $\mathrm{A_u}$), similarly to smaller fullerenes~\cite{Konstantinidis2009} and other frustrated magnetic molecules~\cite{Rousochatzakis2008,Rausch2022SOD60}.
The expectation values of $S^2$ for the GCNN wave functions are listed in Table~\ref{tab: C80 S2}, which deviate from the expected values $S(S+1)$ substantially more than in the C$_{60}$ case. 
This is to be expected, given the much smaller spacing between energy levels (e.g., we find a quintet state in every space-group irrep within an energy window of $0.2J$). 
Nevertheless, all odd-parity states can be identified as predominantly triplet, and most even-parity states as $S=0$ or $S=2$.
An exception is $\mathrm{T_{2g}}, P=+1$, whose $\langle {S}^2\rangle$ is consistent with a $2:1$ mixture of a singlet and a quintet.
Likewise, the $\mathrm{T_{1g}}, S^z=2$ calculation reproducibly converges to $\langle {S}^2\rangle\approx 12$, consistent with $S=3$ rather than $S=2$, even though the $P=+1$ calculation in the same sector converges to a quintet at a lower energy.
In both cases, however, the energy difference between the states in question is extremely small and may not be enough to guide the optimisation algorithm to a perfect ${S}^2$ eigenstate, allowing expressivity limitations of the GCNN ansatz to dominate the optimisation trajectory.
To the best of our knowledge, there are no variational-energy benchmarks for C$_{80}$ to which our results could be compared. 

The most striking features of the spectrum in Fig.~\ref{fig:c80-summary} are the near-degenerate singlet ``ground states'' in the two one-dimensional irreps $\mathrm{A_u,A_g}$,
as well as the presence of triplet and quintet excitations in almost all symmetry sectors within a very narrow energy range.
We will not attempt to account for every state in this dense spectrum, but only highlight the lowest-energy pair of states in each spin sector (circled in Fig.~\ref{fig:c80-summary}), all of which follow the pattern of incipient inversion-symmetry breaking seen for the ground state:
$\mathrm{A_u, A_g}$ for $S=0$;
$\mathrm{T_{2g},T_{2u}}$ for $S=1$; and 
$\mathrm{H_u,H_g}$ for $S=2$.

\begin{figure*}
    \centering
    \includegraphics{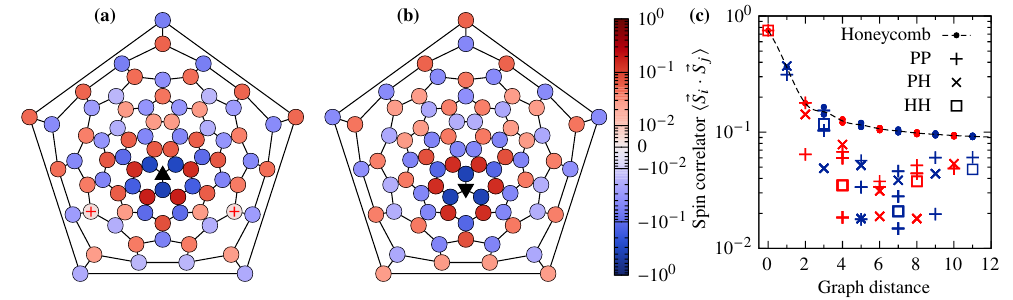}
    \caption{(a,b) Ground-state spin--spin correlation functions $\langle \vec{S}_i\cdot\vec{S}_j\rangle$ in the C$_{80}$ geometry for P-type (a) and H-type (b) reference points $i$ (marked with a black triangle). (c) Spin--spin correlations as a function of graph distance. Red and blue symbols stand for positive and negative correlators, respectively. Coloured dots show the spin correlation functions of a 512-sites honeycomb lattice, measured using QMC; the dashed line is a spline connecting these dots and is included as a guide to the eye. One value below 0.01 in magnitude [$1.90(8)\times10^{-3}$, at graph distance 6, highlighted with coloured symbols in (a)] was truncated for visibility.}
    \label{fig: C80 correlators}
\end{figure*}

Unlike the C$_{60}$ geometry, there are two symmetry-inequivalent kinds of lattice site:
60 sites (labelled P) belong to one of the 12 pentagonal faces;
the remaining 20 (labelled H) are surrounded by three hexagonal faces each.
Fig.~\ref{fig: C80 correlators} shows the spatial structure of spin correlation functions around both kinds of site in the ($\mathrm{A_u}$) ground state. (The correlation structure of the $\mathrm{A_g}$ singlet is visually indistinguishable.)
The sign of correlations again alternates with graph distance, reminiscent of Néel ordering;
unlike C$_{60}$, however, this alternating pattern persists without any anomalies all the way to the antipodal points.
Similar to C$_{60}$, the magnitude of correlators dips at intermediate graph distances before increasing and levelling off for the largest distances at a typical value of about $\pm 0.05$, well above the equivalent figure for C$_{60}$.
This is consistent with the diminishing effect of frustration expected for large fullerenes: 
In the limit of an infinitely large fullerene ``molecule'' (which, however, still has only 12 pentagonal faces), we expect to recover the ground-state behaviour of the honeycomb Heisenberg antiferromagnet, which forms a N\'eel order with spin correlator $\langle \vec{S}_0\cdot \vec{S}_r\rangle = \pm0.0717(3)$~\cite{Castro2006HoneycombQMC} in the long-distance limit [cf.~solid dots in Fig.~\ref{fig: C80 correlators}(c)].

\begin{figure}
    \centering
    \includegraphics{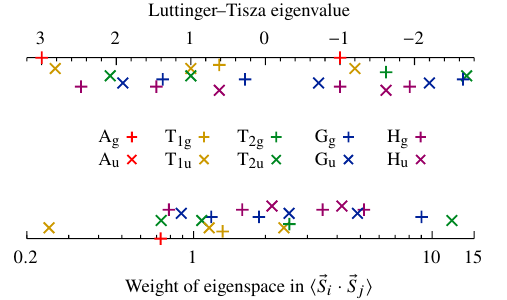}
    \caption{Top: eigenvalues of the classical Hamiltonian matrix (i.e., the adjacency matrix) in the C$_{80}$ geometry. Note that the axis is reversed, so the lowest eigenvalue (the Luttinger--Tisza ground state) is to the right. Bottom: log-scale plot of the weight (eigenvalue times degeneracy) of the eigenspaces of the spin correlator matrix $\langle \vec S_i \cdot \vec S_j\rangle$. The eigenvalue of one $\mathrm{A_g}$ eigenvector, corresponding to the net magnetisation, is zero within Monte Carlo error and is not plotted.}
    \label{fig: C80 LT spectrum}
\end{figure}

These features of the ground state and the low-lying spectrum can again be accounted for in terms of incipient Néel ordering.
The ground state of the large-$S$ Heisenberg Hamiltonian, obtained either by direct simulation or the Luttinger--Tisza method, is again noncoplanar, transforming under the $\mathrm{T_{2u}}$ irrep of $I_h$.
The same irrep is also dominant in the spectrum of the correlator matrix (Fig.~\ref{fig: C80 LT spectrum}); the gap to subleading eigenvalues is increased compared to C$_{60}$, as expected for an incipient Bragg peak.

\begin{figure}
    \centering
    \includegraphics{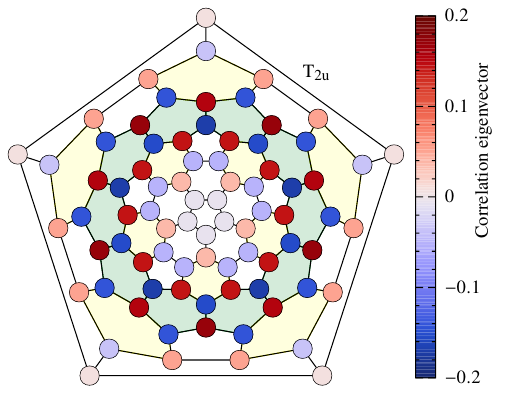}
    \caption{Eigenvector of the ground-state correlator matrix $\langle \vec{S}_i \cdot\vec{S}_j \rangle$ in the C$_{80}$ geometry corresponding to its largest eigenvalue. Green and yellow hexagons indicate a maximal unfrustrated portion of the C$_{80}$ structure.}
    \label{fig: C80 PCA}
\end{figure}

A representative eigenvector from this leading $\mathrm{T_{2u}}$ irrep is plotted in Fig.~\ref{fig: C80 PCA}. 
Similar to C$_{60}$, it forms a Néel pattern on the largest unfrustrated subgraph of the fullerene structure: 
The 10 hexagons (green) on the ``equator'' of the molecule are, however, laid out differently in the two molecules, thus antipodal points in C$_{80}$ acquire opposite signs in the Néel pattern, which explains the inversion-odd irrep.
Furthermore, the unfrustrated support of the Néel pattern includes 10 further hexagons (yellow);
however, these are separated from one another by pentagons, which frustrate and reduce the ordering amplitude.

Similar to C$_{60}$, the low-lying spectrum can be predicted either by applying the tower-of-states formalism to the classical ground state, or by constructing $\mathrm{T_{2u}}$ Goldstone operators~\eqref{eq: magnon operator} from the leading eigenvectors of the correlator matrix. 
The first generates the inversion-broken low-energy spectrum seen numerically:
$\mathrm{A_g, A_u}$ for $S=0$;
$\mathrm{T_{2g},T_{2u}}$ for $S=1$; and 
$\mathrm{H_g,H_u}$ for $S=2$.
On the other hand, acting with the Goldstone operators on the $\mathrm{A_u,A_g}$ ground states once yields $\mathrm{T_{2g},T_{2u}}$ triplets, while the two-Goldstone manifold includes $\mathrm{H_u,H_g}$ quintets~%
\footnote{The Goldstone-operator construction also predicts $\mathrm{A_u,A_g}$ quintets. These are indeed at a similar energy as the $\mathrm{H_u,H_g}$ ones, but are not as clearly separate from the rest of the $S=2$ spectrum as in the C$_{60}$ case. }.
The latter construction also accounts for the energy ordering of the nearly degenerate pairs: those derived from the $\mathrm{A_u}$ ground state ($\mathrm{T_{2g}}$ triplet, $\mathrm{H_u}$ quintet) are all lower in energy than the tower of the $\mathrm{A_g}$ singlet ($\mathrm{T_{2u}}$ triplet, $\mathrm{H_g}$ quintet).

The Goldstone-mode operators also give microscopic account of the apparent inversion-symmetry breaking in the spectrum.
Since the $\mathrm{T_{2u}}$ irrep is threefold degenerate, we can construct three independent triplet operators using~\eqref{eq: magnon operator}, which we label as $\vec{\cal S}_1,\vec{\cal S}_2,\vec{\cal S}_3$. 
Using all three of these, we can uniquely construct the singlet operator $\hat{\cal X} = \vec{\cal S}_1 \cdot (\vec{\cal S}_2 \times \vec{\cal S}_3)$:
one can verify that this operator transforms under the $\mathrm{A_u}$ irrep~\footnote{This can be seen without detailed calculation: the product of three inversion-odd operators is inversion-odd, and the only inversion-odd one-dimensional irrep of $I_h$ is $\mathrm{A_u}$},
thus it might map the $\mathrm{A_g}$ and $\mathrm{A_u}$ singlet ``ground states'' on one another.
Indeed, the overlap of $\mathcal{X}|\mathrm{A_{u(g)}}\rangle$ and $|\mathrm{A_{g(u)}}\rangle$ is 0.743 (0.693), high values for 80-site many-body states.

The operator $\mathcal{X}$ is odd under both inversion and time-reversal symmetry, which strongly suggests that the incipient breaking of inversion symmetry is chiral in nature.
The situation is somewhat similar to that of the tetrahedral order found in the triangular-lattice $J_1-J_2-J_\chi$ model~\cite{wietek2017chiral,Wietek2017TOS}:
its tower of states can be captured in terms of three gapless triplet operators, the triple product of which breaks both mirror and time-reversal symmetry.
This order, however, is only stabilised on the triangular lattice by a substantial $J_\chi$ coupling, which breaks these symmetries explicitly,
while in C$_{80}$, the incipient chiral order emerges spontaneously.

\section{Larger fullerenes}
\label{sec: large fullerene}

\begin{figure}
    \centering
    \includegraphics{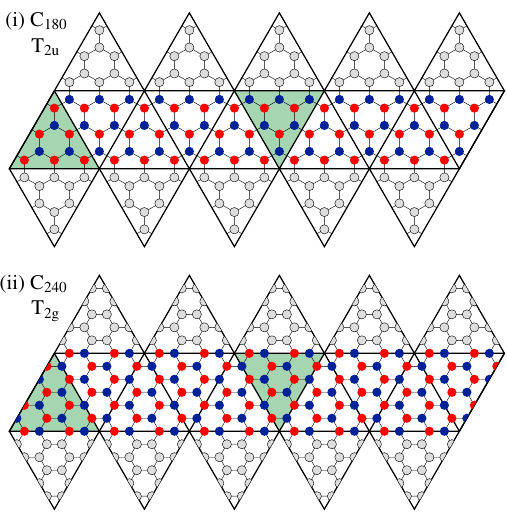}
    \caption{Structure of the ansatz for the dominant eigenvector of $C_{ij}$ on fullerenes of the $I_h$-symmetric (i) C$_{20n^2}$ and (ii) C$_{60n^2}$ series, shown on the flattened net of an icosahedron. They form a Néel pattern of positive (red) and negative (blue) amplitudes on an unfrustrated belt covering 10 of the 20 icosahedron faces, and have only small support further away (gray). Under proper rotations, both transform as the $\mathrm{T_2}$ irrep of $I$. 
    Under inversion, the two green triangles map onto each other; the arrangement of spins transforms as if the triangle was horizontally flipped on the diagram. This flips the sign of the C$_{20n^2}$ structure and preserves that of C$_{60n^2}$. }
    \label{fig:large n}
\end{figure}

It is very natural to ask what determines whether a given fullerene geometry supports such incipient chiral-symmetry breaking.
The mechanism proposed above severely restricts the number of suitable symmetry groups and allotropes,
as requires at least a noncoplanar classical ground state (usually associated with a threefold degenerate ground state in the Luttinger--Tisza spectrum, or a threefold degenerate leading ``Bragg-peak'' eigenvalue in the quantum correlation matrix), as well as inversion symmetry (so it can be broken).
In the following, we explore families of fullerene structures that satisfy these requirements; for a numerical exploration of classical ground-state degeneracy in other geometries, see Ref.~\cite{Konstantinidis2023}.

Let us first focus on fullerenes with full icosahedral ($I_h$) symmetry. These can all be constructed by covering the 20 faces of an icosahedron with patches of the honeycomb lattice~\cite{King2006LargeFullerenes}.
For $I_h$ symmetry, this covering has to be symmetric under every symmetry of a single triangle, which leads to two distinct series of solutions (Fig.~\ref{fig:large n}):
(i) there are $n^2$ lattice points on every face, none of which is shared between more than one face, for a total of $20n^2$ sites;
(ii) some honeycomb edges are aligned with the edges of the icosahedron, yielding a total of $60n^2$ sites.
We conjecture that the incipient Bragg peaks of both kinds of structure are threefold degenerate:
more specifically, we expect that the leading correlation eigenvectors show a pronounced Néel pattern on 10 of 20 faces of the icosahedron, which fades away once frustration occurs, as sketched in Fig.~\ref{fig:large n}.
We can associate a Néel order parameter with each face, which remains well-defined (i.e., does not change sign) under proper rotations; 
therefore, the pattern transform under the same irrep of the chiral icosahedral group $I$ as the classical ground state of the C$_{20}$ Heisenberg model, namely $\mathrm{T_2}$.
Under inversion, however, the two series behave differently:
in the C$_{20n^2}$ case, inversion-related sites have opposite spins (in fact, each triangle has a net magnetisation, which is odd under inversion), while the C$_{60n^2}$ ansatz is inversion-even.
That is, the lowest triplet excitation of all  C$_{20n^2}$ fullerenes is expected to transform as $\mathrm{T_{2u}}$ (leading to incipient chiral order), while  C$_{60n^2}$ molecules have low-lying $\mathrm{T_{2g}}$ triplet excitations.
We verified this prediction by numerically obtaining the classical ground states of the two sequences up to $n=6$ (C$_{720}$) and $n=4$ (C$_{960}$), respectively.

Beyond, icosahedral symmetry, the ingredients proposed above are also available in the largest achiral subgroup of $I_h$, the pyritohedral group $T_h$, which too has three-dimensional irreps $\mathrm{T_g}$ and $\mathrm{T_u}$ (corresponding to both $\mathrm{T_1}$ and $\mathrm{T_2}$ of $I_h$). 
This is particularly important for crystalline compounds of fullerenes, which cannot maintain full icosahedral symmetry.
However, in systems such as the superconductor $\mathrm{K_3C_{60}}$~\cite{C60_superconductivity}, the fullerenes retain full pyritohedral symmetry~\cite{Allen1995_K3C60_Structure}, thus our results remain relevant for their behaviour.
More generally, noncoplanar ground states may arise even if they are composed of more than one irrep of the point group, especially if the classical ground state predicted by the Luttinger--Tisza method has support only on parts of the molecule~%
\footnote{An example is the C$_{70}$ rugbyballene molecule, whose Luttinger--Tisza ground state (in the $\mathrm{A_2'}$ irrep of the $D_{5h}$ point group) vanishes on 20 sites near the ends of the molecule, requiring an additional $\mathrm{E_2'}$ component to form a valid classical ground state. However, this molecule is not inversion symmetric, so no chiral state in the strict sense may form on it.}.
Such classical ground states would still support towers of states with three low-lying spin-triplet excitations, which may combine to generate an operator that breaks inversion and time-reversal symmetry.

\section{Conclusion}
\label{sec: conclusion}

In summary, we demonstrated that competition between the Néel ordering tendency of the honeycomb lattice and frustrated pentagonal faces leads to a feature-rich incipient noncoplanar order in the quantum Heisenberg model on large fullerene lattices.
We generalised a number of techniques commonly used to diagnose ordering tendencies on finite lattices to the molecular case, which lacks translation symmetry:
\begin{enumerate}
    \item Bragg peaks can be defined in real space as diverging eigenvalues of the spin correlator matrix; in a molecule, they are labelled by point-group irreps rather than wave vectors.
    \item Even in the deep quantum limit $S=1/2$, these irreps can be predicted from the ground state of the large-$S$ version of the Hamiltonian, which can be constructed analogously to the Luttinger--Tisza method.
    \item The low-energy spectrum is characterised by a tower of states, the quantum numbers of which can be predicted using ansatz Goldstone-mode operators constructed from the incipient Bragg peaks.
\end{enumerate}
We used these approaches to analyse the low-energy spectrum and ground-state wave function of the nearest-neighbour spin-$1/2$ Heisenberg model on the icosahedrally-symmetric C$_{60}$ and C$_{80}$ fullerene geometries.
Our numerical results were obtained from variational Monte Carlo using group-convolutional neural-network (GCNN) wave-function ansätze, which allowed us to construct the symmetry-resolved low-energy spectrum in detail.
We benchmarked the method against ED on a C$_{32}$ allotrope and DMRG on C$_{60}$, achieving excellent variational energies in both cases.

For buckminsterfullerene, we found an incipient Bragg peak transforming under the $\mathrm{T_{2g}}$ irrep of the icosahedral point group, which allows for the formation of a Néel pattern on the largest unfrustrated subset of the fullerene graph.
This matches the noncoplanar ground state of the classical Heisenberg model~\cite{Coffey1992};
furthermore, the low-energy spectrum of $S=0,1,2$ excitations follows a tower-of-states structure derived from a triplet Goldstone-mode excitation transforming under the same irrep,
pointing towards an incipient noncoplanar order with pronounced Néel-like features on this relatively small and highly frustrated system.

We find similar ordering tendencies on the C$_{80}$ geometry.
The structure of the incipient Bragg peak is again determined by maximally covering the graph with a Néel pattern, which leads to a triplet of leading correlation eigenvectors transforming under the $\mathrm{T_{2u}}$ irrep.
The tower of states corresponding to this inversion-odd incipient order consists of pairs of nearly degenerate (multiplets of) states, distinguished by their inversion eigenvalue.
We can relate the wave functions of these states using an explicit operator constructed from Goldstone modes, which is odd under both inversion and time-reversal, indicating that the incipient ordering in C$_{80}$ is chiral. 
Such inversion-symmetry breaking may have interesting ramifications for optical probes: for instance, by breaking down the rule of mutual exclusion~\cite{bernath2005spectra}, it may make Raman-active modes of C$_{80}$ visible in infrared spectroscopy or vice versa.
Furthermore, the noncoplanar magnetic textures of both molecules may induce anomalous Hall response~\cite{Yamaguchi2021MagneticTextureAHE} that may in turn affect the superconductivity of such materials as $\mathrm{K_3C_{60}}$.
Here too, the chiral magnetic ground state of C$_{80}$ may open the door to more exotic superconducting behaviour, which will be interesting to explore in future theoretical and experimental work.

Finally, we argue that our findings are not limited to the C$_{60}$ and C$_{80}$ geometries, but are relevant for much wider class of fullerene geometries. 
A case in point are the two sequences of fullerenes with full icosahedral symmetry, shown in Fig.~\ref{fig:large n}, whose differing geometries lead to chiral incipient ordering in one sequence but not in the other, a surprisingly persistent frustration effect even in the limit of very large molecules.
Numerically exploring these large molecules, fullerenes with lower symmetry, other magnetic molecules (e.g. with icosidodecahedral symmetry and larger spins~\cite{Neuscamman2012,Ummethum2013}), as well as the intermediate-$U$ Hubbard model more relevant for real fullerenes, are all exciting directions for future work, calling for improvements to our current neural-quantum-state techniques, which will in turn also benefit studies of lattice models. 
From a technical point of view, tensor-network studies of C$_{80}$ would also be interesting, as they may show time-reversal symmetry breaking explicitly~\cite{Cirac2017FundamentalMPS}.

\begin{acknowledgments}
    We thank Alexander Wietek for helpful discussions.
    We are especially grateful to Roman Rausch for providing us with additional DMRG data for C$_{60}$, and to Christopher Roth and Maxime Thumin for their participation at the early stages of this work.
    NQS simulations were performed using the NetKet~\cite{Vicentini2022NetKetSystems} library. 
    Reference stochastic series expansion data for the honeycomb lattice was obtained using the ALPS~\cite{Alps2} library.
    All heat maps use perceptionally uniform colour maps developed in Ref.~\cite{Kovesi2015GoodThem}.
    Computing resources were provided by STFC Scientific Computing Department’s STFC Cloud service.
    This work was granted access to the HPC resources of CALMIP center under the allocation 2022-P0677 as well as GENCI (grant A0130500225). This study has been (partially) supported through the EUR grant NanoX No. ANR-17-EURE-0009 in the framework of the ``Programme des Investissements d'Avenir.'' This work benefited from the support of the project QMAHT ANR-22-CE30-0032-03 of the French National Research Agency (ANR).
    A.~Sz.\ gratefully acknowledges the ISIS Neutron and Muon Source and the Oxford--ShanghaiTech collaboration for support of the Keeley--Rutherford fellowship at Wadham College, Oxford. 
    For the purpose of open access, the authors have applied a Creative Commons Attribution (CC-BY) licence to any author accepted manuscript version arising.
\end{acknowledgments}

\appendix

\section{Projecting on subspaces of higher-dimensional irreps}
\label{app: projection}

As outlined in Sec.~\ref{sec: irrep projection}, we restricted ansätze transforming under higher-dimensional irreps of the space group $G$ onto a one-dimensional subspace of the symmetry-protected multiplets by imposing additional symmetry constraints, as follows.
Consider an Abelian subgroup $H$ of the space group $G$. 
Restricting any irrep $\chi$ of $G$ onto $H$ gives a valid representation thereof; for multi-dimensional irreps, however, this is no longer an irrep of $H$, but can be decomposed into them.
If an irrep $\chi_H$ of $H$ appears with multiplicity 1 in this decomposition, we can select a unique representative of $\chi$ by first projecting onto $\chi_H$ using~\eqref{eq: irrep projection formula}, then onto $\chi$ itself:
\begin{align}
    |\psi\rangle &= 
    \frac{d_\chi}{|G|}\sum_{g\in G} \chi_g^* \hat g  \;
    \frac{1}{|H|}\sum_{h\in H} \chi_{H,h}^* \hat h  \;
    |\psi_0\rangle \nonumber\\
    &= \frac{d_\chi}{|G||H|} \sum_{g\in G} \bigg(\sum_{h\in H} \chi_{gh^{-1}} \chi_{H,h}\bigg)^* \hat g\; |\psi_0\rangle.
\end{align}
This has the same form as the original projection~\eqref{eq: irrep projection formula}, with the ``effective character''
\begin{equation}
    \tilde\chi_g = \frac1{|H|}\sum_{h\in H} \chi_{gh^{-1}}\chi_{H,h}.
    \label{eq: effective character}
\end{equation}
For simplicity, we choose $H$ for each irrep such that their trivial irrep appears with multiplicity 1, so Eq.~\eqref{eq: effective character} reduces to averaging characters in (right) cosets of $H$.

The $D_{3h}$ point group of C$_{32}$ has two two-dimensional irreps, $\mathrm{E'}$ and $\mathrm{E''}$. A suitable choice of $H$ for both is the $C_2$ subgroup generated by one of the three $180^\circ$ rotations.

The $I_h$ point group of C$_{60}$ and C$_{80}$ has three-, four-, and five-dimensional irreps. We decomposed each of these using the following $H$:
\begin{itemize}
    \item T irreps: any $C_3$ rotation subgroup
    \item G irreps: the $C_{2v}$ subgroup that fixes an edge of a dodecahedron with the same symmetry group. In C$_{60}$, the same group fixes an edge lying between two hexagons. In C$_{80}$, it fixes the hexagon diagonal lying between two nearest-neighbour pentagons.
    \item H irreps: any $C_5$ rotation subgroup.
\end{itemize}
Each of these groups decompose the given irrep into every one of their own irreps, each with multiplicity 1.

Figures~\ref{fig: C60 PCA} and~\ref{fig: C80 PCA} are exceptions:
there, we decomposed irreps of both the correlation eigenvectors and the low-lying wave functions into irreps of the $C_5$ rotation group around the centre of the Schlegel plot.
Such a decomposition of T irreps contains the trivial irrep, yielding rotationally symmetric plots.
The $\mathrm{G_u}$ irrep in Fig.~\ref{fig: C60 PCA}(b), by contrast, decomposes into the four nontrivial irreps of $C_5$: for the plot, we used the real part of the $e^{2\pi i/5}$ rotation-eigenvalue component (that is, a linear combination of the $e^{\pm2\pi i/5}$ components).

\section{Tables of optimal variational energies}
\label{app:data}

GCNN variational energies in every symmetry sector we considered are listed for C$_{32}$, C$_{60}$, and C$_{80}$ in Tables~\ref{tab: C32 energy}, \ref{tab: C60 energy}, and \ref{tab: C80 energy}, respectively.
In addition, we performed exact diagonalisation on the C$_{32}$ Heisenberg model to extend the results in Ref.~\cite{Konstantinidis2009} to every eigenstate below energy $-15J$; these energies are listed in Table~\ref{tab: C32 ED}. 

\bibliography{fullerene,attila_everything,etc}

\begin{thebibliography}{61}%
\makeatletter
\providecommand \@ifxundefined [1]{%
 \@ifx{#1\undefined}
}%
\providecommand \@ifnum [1]{%
 \ifnum #1\expandafter \@firstoftwo
 \else \expandafter \@secondoftwo
 \fi
}%
\providecommand \@ifx [1]{%
 \ifx #1\expandafter \@firstoftwo
 \else \expandafter \@secondoftwo
 \fi
}%
\providecommand \natexlab [1]{#1}%
\providecommand \enquote  [1]{``#1''}%
\providecommand \bibnamefont  [1]{#1}%
\providecommand \bibfnamefont [1]{#1}%
\providecommand \citenamefont [1]{#1}%
\providecommand \href@noop [0]{\@secondoftwo}%
\providecommand \href [0]{\begingroup \@sanitize@url \@href}%
\providecommand \@href[1]{\@@startlink{#1}\@@href}%
\providecommand \@@href[1]{\endgroup#1\@@endlink}%
\providecommand \@sanitize@url [0]{\catcode `\\12\catcode `\$12\catcode
  `\&12\catcode `\#12\catcode `\^12\catcode `\_12\catcode `\%12\relax}%
\providecommand \@@startlink[1]{}%
\providecommand \@@endlink[0]{}%
\providecommand \url  [0]{\begingroup\@sanitize@url \@url }%
\providecommand \@url [1]{\endgroup\@href {#1}{\urlprefix }}%
\providecommand \urlprefix  [0]{URL }%
\providecommand \Eprint [0]{\href }%
\providecommand \doibase [0]{https://doi.org/}%
\providecommand \selectlanguage [0]{\@gobble}%
\providecommand \bibinfo  [0]{\@secondoftwo}%
\providecommand \bibfield  [0]{\@secondoftwo}%
\providecommand \translation [1]{[#1]}%
\providecommand \BibitemOpen [0]{}%
\providecommand \bibitemStop [0]{}%
\providecommand \bibitemNoStop [0]{.\EOS\space}%
\providecommand \EOS [0]{\spacefactor3000\relax}%
\providecommand \BibitemShut  [1]{\csname bibitem#1\endcsname}%
\let\auto@bib@innerbib\@empty
\bibitem [{\citenamefont {Anderson}(1952)}]{Anderson1952TOS}%
  \BibitemOpen
  \bibfield  {author} {\bibinfo {author} {\bibfnamefont {P.~W.}\ \bibnamefont
  {Anderson}},\ }\bibfield  {title} {\bibinfo {title} {An approximate quantum
  theory of the antiferromagnetic ground state},\ }\href
  {https://doi.org/10.1103/PhysRev.86.694} {\bibfield  {journal} {\bibinfo
  {journal} {Phys. Rev.}\ }\textbf {\bibinfo {volume} {86}},\ \bibinfo {pages}
  {694} (\bibinfo {year} {1952})}\BibitemShut {NoStop}%
\bibitem [{\citenamefont {Dyson}\ \emph {et~al.}(1978)\citenamefont {Dyson},
  \citenamefont {Lieb},\ and\ \citenamefont {Simon}}]{Dyson1978}%
  \BibitemOpen
  \bibfield  {author} {\bibinfo {author} {\bibfnamefont {F.~J.}\ \bibnamefont
  {Dyson}}, \bibinfo {author} {\bibfnamefont {E.~H.}\ \bibnamefont {Lieb}},\
  and\ \bibinfo {author} {\bibfnamefont {B.}~\bibnamefont {Simon}},\ }\bibfield
   {title} {\bibinfo {title} {Phase transitions in quantum spin systems with
  isotropic and nonisotropic interactions},\ }\href
  {https://doi.org/10.1007/BF01106729} {\bibfield  {journal} {\bibinfo
  {journal} {J. Stat. Phys.}\ }\textbf {\bibinfo {volume} {18}},\ \bibinfo
  {pages} {335} (\bibinfo {year} {1978})}\BibitemShut {NoStop}%
\bibitem [{\citenamefont {Kennedy}\ \emph {et~al.}(1988)\citenamefont
  {Kennedy}, \citenamefont {Lieb},\ and\ \citenamefont
  {Shastry}}]{Kennedy1988}%
  \BibitemOpen
  \bibfield  {author} {\bibinfo {author} {\bibfnamefont {T.}~\bibnamefont
  {Kennedy}}, \bibinfo {author} {\bibfnamefont {E.~H.}\ \bibnamefont {Lieb}},\
  and\ \bibinfo {author} {\bibfnamefont {B.~S.}\ \bibnamefont {Shastry}},\
  }\bibfield  {title} {\bibinfo {title} {Existence of {N\'eel} order in some
  spin-1/2 {Heisenberg} antiferromagnets},\ }\href
  {https://doi.org/10.1007/BF01023854} {\bibfield  {journal} {\bibinfo
  {journal} {J. Stat. Phys.}\ }\textbf {\bibinfo {volume} {53}},\ \bibinfo
  {pages} {1019} (\bibinfo {year} {1988})}\BibitemShut {NoStop}%
\bibitem [{\citenamefont {Neves}\ and\ \citenamefont
  {Perez}(1986)}]{Neves1986}%
  \BibitemOpen
  \bibfield  {author} {\bibinfo {author} {\bibfnamefont {E.~J.}\ \bibnamefont
  {Neves}}\ and\ \bibinfo {author} {\bibfnamefont {J.}~\bibnamefont {Perez}},\
  }\bibfield  {title} {\bibinfo {title} {Long range order in the ground state
  of two-dimensional antiferromagnets},\ }\href
  {https://doi.org/https://doi.org/10.1016/0375-9601(86)90571-2} {\bibfield
  {journal} {\bibinfo  {journal} {Phys. Lett. A}\ }\textbf {\bibinfo {volume}
  {114}},\ \bibinfo {pages} {331} (\bibinfo {year} {1986})}\BibitemShut
  {NoStop}%
\bibitem [{\citenamefont {Wietek}\ \emph {et~al.}(2017)\citenamefont {Wietek},
  \citenamefont {Schuler},\ and\ \citenamefont {L\"auchli}}]{Wietek2017TOS}%
  \BibitemOpen
  \bibfield  {author} {\bibinfo {author} {\bibfnamefont {A.}~\bibnamefont
  {Wietek}}, \bibinfo {author} {\bibfnamefont {M.}~\bibnamefont {Schuler}},\
  and\ \bibinfo {author} {\bibfnamefont {A.~M.}\ \bibnamefont {L\"auchli}},\
  }\href@noop {} {\bibinfo {title} {Studying continuous symmetry breaking using
  energy level spectroscopy}} (\bibinfo {year} {2017}),\ \Eprint
  {https://arxiv.org/abs/1704.08622} {arXiv:1704.08622 [cond-mat.str-el]}
  \BibitemShut {NoStop}%
\bibitem [{\citenamefont {Wietek}\ and\ \citenamefont
  {L\"auchli}(2017)}]{wietek2017chiral}%
  \BibitemOpen
  \bibfield  {author} {\bibinfo {author} {\bibfnamefont {A.}~\bibnamefont
  {Wietek}}\ and\ \bibinfo {author} {\bibfnamefont {A.~M.}\ \bibnamefont
  {L\"auchli}},\ }\bibfield  {title} {\bibinfo {title} {Chiral spin liquid and
  quantum criticality in extended $s=\frac{1}{2}$ heisenberg models on the
  triangular lattice},\ }\href {https://doi.org/10.1103/PhysRevB.95.035141}
  {\bibfield  {journal} {\bibinfo  {journal} {Phys. Rev. B}\ }\textbf {\bibinfo
  {volume} {95}},\ \bibinfo {pages} {035141} (\bibinfo {year}
  {2017})}\BibitemShut {NoStop}%
\bibitem [{\citenamefont {Wietek}\ and\ \citenamefont
  {L\"auchli}(2020)}]{Wietek2020KagomeVBS}%
  \BibitemOpen
  \bibfield  {author} {\bibinfo {author} {\bibfnamefont {A.}~\bibnamefont
  {Wietek}}\ and\ \bibinfo {author} {\bibfnamefont {A.~M.}\ \bibnamefont
  {L\"auchli}},\ }\bibfield  {title} {\bibinfo {title} {{Valence bond solid and
  possible deconfined quantum criticality in an extended kagome lattice
  Heisenberg antiferromagnet}},\ }\href
  {https://doi.org/10.1103/PhysRevB.102.020411} {\bibfield  {journal} {\bibinfo
   {journal} {Phys. Rev. B}\ }\textbf {\bibinfo {volume} {102}},\ \bibinfo
  {pages} {020411} (\bibinfo {year} {2020})}\BibitemShut {NoStop}%
\bibitem [{\citenamefont {Wietek}\ \emph {et~al.}(2023)\citenamefont {Wietek},
  \citenamefont {Capponi},\ and\ \citenamefont
  {Läuchli}}]{wietek2023triangular}%
  \BibitemOpen
  \bibfield  {author} {\bibinfo {author} {\bibfnamefont {A.}~\bibnamefont
  {Wietek}}, \bibinfo {author} {\bibfnamefont {S.}~\bibnamefont {Capponi}},\
  and\ \bibinfo {author} {\bibfnamefont {A.~M.}\ \bibnamefont {Läuchli}},\
  }\href@noop {} {\bibinfo {title} {Quantum electrodynamics in 2+1 dimensions
  as the organizing principle of a triangular lattice antiferromagnet}}
  (\bibinfo {year} {2023}),\ \Eprint {https://arxiv.org/abs/2303.01585}
  {arXiv:2303.01585 [cond-mat.str-el]} \BibitemShut {NoStop}%
\bibitem [{\citenamefont {Nomura}\ and\ \citenamefont
  {Imada}(2021)}]{Nomura2021Dirac-TypeSpectroscopy}%
  \BibitemOpen
  \bibfield  {author} {\bibinfo {author} {\bibfnamefont {Y.}~\bibnamefont
  {Nomura}}\ and\ \bibinfo {author} {\bibfnamefont {M.}~\bibnamefont {Imada}},\
  }\bibfield  {title} {\bibinfo {title} {{Dirac-Type Nodal Spin Liquid Revealed
  by Refined Quantum Many-Body Solver Using Neural-Network Wave Function,
  Correlation Ratio, and Level Spectroscopy}},\ }\href
  {https://doi.org/10.1103/PHYSREVX.11.031034} {\bibfield  {journal} {\bibinfo
  {journal} {Phys. Rev. X}\ }\textbf {\bibinfo {volume} {11}},\ \bibinfo
  {pages} {031034} (\bibinfo {year} {2021})}\BibitemShut {NoStop}%
\bibitem [{\citenamefont {Wulferding}\ \emph {et~al.}(2019)\citenamefont
  {Wulferding}, \citenamefont {Choi}, \citenamefont {Lee},\ and\ \citenamefont
  {Choi}}]{Wulferding2020}%
  \BibitemOpen
  \bibfield  {author} {\bibinfo {author} {\bibfnamefont {D.}~\bibnamefont
  {Wulferding}}, \bibinfo {author} {\bibfnamefont {Y.}~\bibnamefont {Choi}},
  \bibinfo {author} {\bibfnamefont {W.}~\bibnamefont {Lee}},\ and\ \bibinfo
  {author} {\bibfnamefont {K.-Y.}\ \bibnamefont {Choi}},\ }\bibfield  {title}
  {\bibinfo {title} {Raman spectroscopic diagnostic of quantum spin liquids},\
  }\href {https://doi.org/10.1088/1361-648X/ab45c4} {\bibfield  {journal}
  {\bibinfo  {journal} {J. Phys. Condens. Matter}\ }\textbf {\bibinfo {volume}
  {32}},\ \bibinfo {pages} {043001} (\bibinfo {year} {2019})}\BibitemShut
  {NoStop}%
\bibitem [{\citenamefont {Schnack}(2010)}]{Schnack2010}%
  \BibitemOpen
  \bibfield  {author} {\bibinfo {author} {\bibfnamefont {J.}~\bibnamefont
  {Schnack}},\ }\bibfield  {title} {\bibinfo {title} {{Effects of frustration
  on magnetic molecules: a survey from Olivier Kahn until today}},\ }\href
  {https://doi.org/10.1039/B925358K} {\bibfield  {journal} {\bibinfo  {journal}
  {Dalton Trans.}\ }\textbf {\bibinfo {volume} {39}},\ \bibinfo {pages} {4677}
  (\bibinfo {year} {2010})}\BibitemShut {NoStop}%
\bibitem [{\citenamefont {Rousochatzakis}\ \emph {et~al.}(2008)\citenamefont
  {Rousochatzakis}, \citenamefont {L\"auchli},\ and\ \citenamefont
  {Mila}}]{Rousochatzakis2008}%
  \BibitemOpen
  \bibfield  {author} {\bibinfo {author} {\bibfnamefont {I.}~\bibnamefont
  {Rousochatzakis}}, \bibinfo {author} {\bibfnamefont {A.~M.}\ \bibnamefont
  {L\"auchli}},\ and\ \bibinfo {author} {\bibfnamefont {F.}~\bibnamefont
  {Mila}},\ }\bibfield  {title} {\bibinfo {title} {Highly frustrated magnetic
  clusters: The kagom\'e on a sphere},\ }\href
  {https://doi.org/10.1103/PhysRevB.77.094420} {\bibfield  {journal} {\bibinfo
  {journal} {Phys. Rev. B}\ }\textbf {\bibinfo {volume} {77}},\ \bibinfo
  {pages} {094420} (\bibinfo {year} {2008})}\BibitemShut {NoStop}%
\bibitem [{\citenamefont {Furrer}\ and\ \citenamefont
  {Waldmann}(2013)}]{Furrer2013}%
  \BibitemOpen
  \bibfield  {author} {\bibinfo {author} {\bibfnamefont {A.}~\bibnamefont
  {Furrer}}\ and\ \bibinfo {author} {\bibfnamefont {O.}~\bibnamefont
  {Waldmann}},\ }\bibfield  {title} {\bibinfo {title} {Magnetic cluster
  excitations},\ }\href {https://doi.org/10.1103/RevModPhys.85.367} {\bibfield
  {journal} {\bibinfo  {journal} {Rev. Mod. Phys.}\ }\textbf {\bibinfo {volume}
  {85}},\ \bibinfo {pages} {367} (\bibinfo {year} {2013})}\BibitemShut
  {NoStop}%
\bibitem [{\citenamefont {Stace}\ and\ \citenamefont
  {O'Brien}(2016)}]{review_fullerene}%
  \BibitemOpen
  \bibfield  {author} {\bibinfo {author} {\bibfnamefont {A.~J.}\ \bibnamefont
  {Stace}}\ and\ \bibinfo {author} {\bibfnamefont {P.}~\bibnamefont
  {O'Brien}},\ }\bibfield  {title} {\bibinfo {title} {{Fullerenes: past,
  present and future, celebrating the 30th anniversary of Buckminster
  Fullerene}},\ }\href {https://doi.org/10.1098/rsta.2016.0278} {\bibfield
  {journal} {\bibinfo  {journal} {Philos. Trans. R. Soc. A}\ }\textbf {\bibinfo
  {volume} {374}},\ \bibinfo {pages} {20160278} (\bibinfo {year}
  {2016})}\BibitemShut {NoStop}%
\bibitem [{\citenamefont {Lin}\ \emph {et~al.}(2007)\citenamefont {Lin},
  \citenamefont {S\o{}rensen}, \citenamefont {Kallin},\ and\ \citenamefont
  {Berlinsky}}]{Lin2007b}%
  \BibitemOpen
  \bibfield  {author} {\bibinfo {author} {\bibfnamefont {F.}~\bibnamefont
  {Lin}}, \bibinfo {author} {\bibfnamefont {E.~S.}\ \bibnamefont
  {S\o{}rensen}}, \bibinfo {author} {\bibfnamefont {C.}~\bibnamefont
  {Kallin}},\ and\ \bibinfo {author} {\bibfnamefont {A.~J.}\ \bibnamefont
  {Berlinsky}},\ }\bibfield  {title} {\bibinfo {title} {{Strong correlation
  effects in the fullerene ${\mathrm{C}}_{20}$ studied using a one-band Hubbard
  model}},\ }\href {https://doi.org/10.1103/PhysRevB.76.033414} {\bibfield
  {journal} {\bibinfo  {journal} {Phys. Rev. B}\ }\textbf {\bibinfo {volume}
  {76}},\ \bibinfo {pages} {033414} (\bibinfo {year} {2007})}\BibitemShut
  {NoStop}%
\bibitem [{\citenamefont {Konstantinidis}(2005)}]{Konstantinidis2005}%
  \BibitemOpen
  \bibfield  {author} {\bibinfo {author} {\bibfnamefont {N.~P.}\ \bibnamefont
  {Konstantinidis}},\ }\bibfield  {title} {\bibinfo {title} {Antiferromagnetic
  {H}eisenberg model on clusters with icosahedral symmetry},\ }\href
  {https://doi.org/10.1103/PhysRevB.72.064453} {\bibfield  {journal} {\bibinfo
  {journal} {Phys. Rev. B}\ }\textbf {\bibinfo {volume} {72}},\ \bibinfo
  {pages} {064453} (\bibinfo {year} {2005})}\BibitemShut {NoStop}%
\bibitem [{\citenamefont {Chakravarty}\ \emph {et~al.}(1991)\citenamefont
  {Chakravarty}, \citenamefont {Gelfand},\ and\ \citenamefont
  {Kivelson}}]{Chakravarty1991C60Hubbard}%
  \BibitemOpen
  \bibfield  {author} {\bibinfo {author} {\bibfnamefont {S.}~\bibnamefont
  {Chakravarty}}, \bibinfo {author} {\bibfnamefont {M.~P.}\ \bibnamefont
  {Gelfand}},\ and\ \bibinfo {author} {\bibfnamefont {S.}~\bibnamefont
  {Kivelson}},\ }\bibfield  {title} {\bibinfo {title} {Electronic correlation
  effects and superconductivity in doped fullerenes},\ }\href
  {https://doi.org/10.1126/science.254.5034.970} {\bibfield  {journal}
  {\bibinfo  {journal} {Science}\ }\textbf {\bibinfo {volume} {254}},\ \bibinfo
  {pages} {970} (\bibinfo {year} {1991})}\BibitemShut {NoStop}%
\bibitem [{\citenamefont {Coffey}\ and\ \citenamefont
  {Trugman}(1992)}]{Coffey1992}%
  \BibitemOpen
  \bibfield  {author} {\bibinfo {author} {\bibfnamefont {D.}~\bibnamefont
  {Coffey}}\ and\ \bibinfo {author} {\bibfnamefont {S.~A.}\ \bibnamefont
  {Trugman}},\ }\bibfield  {title} {\bibinfo {title} {{Magnetic properties of
  undoped ${\mathrm{C}}_{60}$}},\ }\href
  {https://doi.org/10.1103/PhysRevLett.69.176} {\bibfield  {journal} {\bibinfo
  {journal} {Phys. Rev. Lett.}\ }\textbf {\bibinfo {volume} {69}},\ \bibinfo
  {pages} {176} (\bibinfo {year} {1992})}\BibitemShut {NoStop}%
\bibitem [{\citenamefont {Scalettar}\ \emph {et~al.}(1993)\citenamefont
  {Scalettar}, \citenamefont {Moreo}, \citenamefont {Dagotto}, \citenamefont
  {Bergomi}, \citenamefont {Jolicoeur},\ and\ \citenamefont
  {Monien}}]{Scalettar1993}%
  \BibitemOpen
  \bibfield  {author} {\bibinfo {author} {\bibfnamefont {R.~T.}\ \bibnamefont
  {Scalettar}}, \bibinfo {author} {\bibfnamefont {A.}~\bibnamefont {Moreo}},
  \bibinfo {author} {\bibfnamefont {E.}~\bibnamefont {Dagotto}}, \bibinfo
  {author} {\bibfnamefont {L.}~\bibnamefont {Bergomi}}, \bibinfo {author}
  {\bibfnamefont {T.}~\bibnamefont {Jolicoeur}},\ and\ \bibinfo {author}
  {\bibfnamefont {H.}~\bibnamefont {Monien}},\ }\bibfield  {title} {\bibinfo
  {title} {{Ground-state properties of the Hubbard model on a
  ${\mathrm{C}}_{60}$ cluster}},\ }\href
  {https://doi.org/10.1103/PhysRevB.47.12316} {\bibfield  {journal} {\bibinfo
  {journal} {Phys. Rev. B}\ }\textbf {\bibinfo {volume} {47}},\ \bibinfo
  {pages} {12316} (\bibinfo {year} {1993})}\BibitemShut {NoStop}%
\bibitem [{\citenamefont {Konstantinidis}(2009)}]{Konstantinidis2009}%
  \BibitemOpen
  \bibfield  {author} {\bibinfo {author} {\bibfnamefont {N.~P.}\ \bibnamefont
  {Konstantinidis}},\ }\bibfield  {title} {\bibinfo {title} {$s=\frac{1}{2}$
  antiferromagnetic {H}eisenberg model on fullerene-type symmetry clusters},\
  }\href {https://doi.org/10.1103/PhysRevB.80.134427} {\bibfield  {journal}
  {\bibinfo  {journal} {Phys. Rev. B}\ }\textbf {\bibinfo {volume} {80}},\
  \bibinfo {pages} {134427} (\bibinfo {year} {2009})}\BibitemShut {NoStop}%
\bibitem [{\citenamefont {Modine}\ and\ \citenamefont
  {Kaxiras}(1996)}]{Modine1996}%
  \BibitemOpen
  \bibfield  {author} {\bibinfo {author} {\bibfnamefont {N.~A.}\ \bibnamefont
  {Modine}}\ and\ \bibinfo {author} {\bibfnamefont {E.}~\bibnamefont
  {Kaxiras}},\ }\bibfield  {title} {\bibinfo {title} {{Variational
  Hilbert-space-truncation approach to quantum Heisenberg antiferromagnets on
  frustrated clusters}},\ }\href {https://doi.org/10.1103/PhysRevB.53.2546}
  {\bibfield  {journal} {\bibinfo  {journal} {Phys. Rev. B}\ }\textbf {\bibinfo
  {volume} {53}},\ \bibinfo {pages} {2546} (\bibinfo {year}
  {1996})}\BibitemShut {NoStop}%
\bibitem [{\citenamefont {Konstantinidis}(2018)}]{Konstantinidis2018LargerED}%
  \BibitemOpen
  \bibfield  {author} {\bibinfo {author} {\bibfnamefont {N.}~\bibnamefont
  {Konstantinidis}},\ }\bibfield  {title} {\bibinfo {title} {Zero-temperature
  magnetic response of small fullerene molecules at the classical and full
  quantum limit},\ }\href
  {https://doi.org/https://doi.org/10.1016/j.jmmm.2017.09.020} {\bibfield
  {journal} {\bibinfo  {journal} {J. Magn. Magn. Mater.}\ }\textbf {\bibinfo
  {volume} {449}},\ \bibinfo {pages} {55} (\bibinfo {year} {2018})}\BibitemShut
  {NoStop}%
\bibitem [{\citenamefont {Momma}\ and\ \citenamefont
  {Izumi}(2011)}]{Momma2011VESTAData}%
  \BibitemOpen
  \bibfield  {author} {\bibinfo {author} {\bibfnamefont {K.}~\bibnamefont
  {Momma}}\ and\ \bibinfo {author} {\bibfnamefont {F.}~\bibnamefont {Izumi}},\
  }\bibfield  {title} {\bibinfo {title} {{VESTA 3 for three-dimensional
  visualization of crystal, volumetric and morphology data}},\ }\href
  {https://doi.org/10.1107/S0021889811038970} {\bibfield  {journal} {\bibinfo
  {journal} {J. Appl. Crystallogr.}\ }\textbf {\bibinfo {volume} {44}},\
  \bibinfo {pages} {1272} (\bibinfo {year} {2011})}\BibitemShut {NoStop}%
\bibitem [{\citenamefont {Tománek}(2014)}]{Tomanek2014NanocarbonJungle}%
  \BibitemOpen
  \bibfield  {author} {\bibinfo {author} {\bibfnamefont {D.}~\bibnamefont
  {Tománek}},\ }\href {https://doi.org/10.1088/978-1-627-05273-3} {\emph
  {\bibinfo {title} {Guide Through the Nanocarbon Jungle}}},\ 2053-2571\
  (\bibinfo  {publisher} {Morgan \& Claypool Publishers},\ \bibinfo {year}
  {2014})\BibitemShut {NoStop}%
\bibitem [{\citenamefont {Benjamin}\ \emph {et~al.}(2006)\citenamefont
  {Benjamin}, \citenamefont {Ardavan}, \citenamefont {Briggs}, \citenamefont
  {Britz}, \citenamefont {Gunlycke}, \citenamefont {Jefferson}, \citenamefont
  {Jones}, \citenamefont {Leigh}, \citenamefont {Lovett}, \citenamefont
  {Khlobystov}, \citenamefont {Lyon}, \citenamefont {Morton}, \citenamefont
  {Porfyrakis}, \citenamefont {Sambrook},\ and\ \citenamefont
  {Tyryshkin}}]{Benjamin2006}%
  \BibitemOpen
  \bibfield  {author} {\bibinfo {author} {\bibfnamefont {S.~C.}\ \bibnamefont
  {Benjamin}}, \bibinfo {author} {\bibfnamefont {A.}~\bibnamefont {Ardavan}},
  \bibinfo {author} {\bibfnamefont {G.~A.~D.}\ \bibnamefont {Briggs}}, \bibinfo
  {author} {\bibfnamefont {D.~A.}\ \bibnamefont {Britz}}, \bibinfo {author}
  {\bibfnamefont {D.}~\bibnamefont {Gunlycke}}, \bibinfo {author}
  {\bibfnamefont {J.}~\bibnamefont {Jefferson}}, \bibinfo {author}
  {\bibfnamefont {M.~A.~G.}\ \bibnamefont {Jones}}, \bibinfo {author}
  {\bibfnamefont {D.~F.}\ \bibnamefont {Leigh}}, \bibinfo {author}
  {\bibfnamefont {B.~W.}\ \bibnamefont {Lovett}}, \bibinfo {author}
  {\bibfnamefont {A.~N.}\ \bibnamefont {Khlobystov}}, \bibinfo {author}
  {\bibfnamefont {S.~A.}\ \bibnamefont {Lyon}}, \bibinfo {author}
  {\bibfnamefont {J.~J.~L.}\ \bibnamefont {Morton}}, \bibinfo {author}
  {\bibfnamefont {K.}~\bibnamefont {Porfyrakis}}, \bibinfo {author}
  {\bibfnamefont {M.~R.}\ \bibnamefont {Sambrook}},\ and\ \bibinfo {author}
  {\bibfnamefont {A.~M.}\ \bibnamefont {Tyryshkin}},\ }\bibfield  {title}
  {\bibinfo {title} {Towards a fullerene-based quantum computer},\ }\href
  {https://doi.org/10.1088/0953-8984/18/21/S12} {\bibfield  {journal} {\bibinfo
   {journal} {J. Phys. Condens. Matter}\ }\textbf {\bibinfo {volume} {18}},\
  \bibinfo {pages} {S867} (\bibinfo {year} {2006})}\BibitemShut {NoStop}%
\bibitem [{\citenamefont {Hebard}\ \emph {et~al.}(1991)\citenamefont {Hebard},
  \citenamefont {Rosseinsky}, \citenamefont {Haddon}, \citenamefont {Murphy},
  \citenamefont {Glarum}, \citenamefont {Palstra}, \citenamefont {Ramirez},\
  and\ \citenamefont {Kortan}}]{C60_superconductivity}%
  \BibitemOpen
  \bibfield  {author} {\bibinfo {author} {\bibfnamefont {A.~F.}\ \bibnamefont
  {Hebard}}, \bibinfo {author} {\bibfnamefont {M.~J.}\ \bibnamefont
  {Rosseinsky}}, \bibinfo {author} {\bibfnamefont {R.~C.}\ \bibnamefont
  {Haddon}}, \bibinfo {author} {\bibfnamefont {D.~W.}\ \bibnamefont {Murphy}},
  \bibinfo {author} {\bibfnamefont {S.~H.}\ \bibnamefont {Glarum}}, \bibinfo
  {author} {\bibfnamefont {T.~T.~M.}\ \bibnamefont {Palstra}}, \bibinfo
  {author} {\bibfnamefont {A.~P.}\ \bibnamefont {Ramirez}},\ and\ \bibinfo
  {author} {\bibfnamefont {A.~R.}\ \bibnamefont {Kortan}},\ }\bibfield  {title}
  {\bibinfo {title} {Superconductivity at 18 {K} in potassium-doped
  {C${60}$}},\ }\href {https://doi.org/10.1038/350600a0} {\bibfield  {journal}
  {\bibinfo  {journal} {Nature}\ }\textbf {\bibinfo {volume} {350}},\ \bibinfo
  {pages} {600} (\bibinfo {year} {1991})}\BibitemShut {NoStop}%
\bibitem [{\citenamefont {Rausch}\ \emph {et~al.}(2021)\citenamefont {Rausch},
  \citenamefont {Plorin},\ and\ \citenamefont {Peschke}}]{Rausch2021}%
  \BibitemOpen
  \bibfield  {author} {\bibinfo {author} {\bibfnamefont {R.}~\bibnamefont
  {Rausch}}, \bibinfo {author} {\bibfnamefont {C.}~\bibnamefont {Plorin}},\
  and\ \bibinfo {author} {\bibfnamefont {M.}~\bibnamefont {Peschke}},\
  }\bibfield  {title} {\bibinfo {title} {{The antiferromagnetic $S=1/2$
  Heisenberg model on the C$_{60}$ fullerene geometry}},\ }\href
  {https://doi.org/10.21468/SciPostPhys.10.4.087} {\bibfield  {journal}
  {\bibinfo  {journal} {SciPost Phys.}\ }\textbf {\bibinfo {volume} {10}},\
  \bibinfo {pages} {87} (\bibinfo {year} {2021})}\BibitemShut {NoStop}%
\bibitem [{\citenamefont {Roth}\ and\ \citenamefont
  {MacDonald}(2021)}]{Roth2021GroupAccuracy}%
  \BibitemOpen
  \bibfield  {author} {\bibinfo {author} {\bibfnamefont {C.}~\bibnamefont
  {Roth}}\ and\ \bibinfo {author} {\bibfnamefont {A.~H.}\ \bibnamefont
  {MacDonald}},\ }\bibfield  {title} {\bibinfo {title} {{Group Convolutional
  Neural Networks Improve Quantum State Accuracy}},\ }\href
  {https://arxiv.org/abs/2104.05085v3} {\bibfield  {journal} {\bibinfo
  {journal} {arXiv:2104.05085}\ } (\bibinfo {year} {2021})}\BibitemShut
  {NoStop}%
\bibitem [{\citenamefont {Roth}\ \emph {et~al.}(2023)\citenamefont {Roth},
  \citenamefont {Szab\'o},\ and\ \citenamefont
  {MacDonald}}]{Roth2023HighAccuracy}%
  \BibitemOpen
  \bibfield  {author} {\bibinfo {author} {\bibfnamefont {C.}~\bibnamefont
  {Roth}}, \bibinfo {author} {\bibfnamefont {A.}~\bibnamefont {Szab\'o}},\ and\
  \bibinfo {author} {\bibfnamefont {A.~H.}\ \bibnamefont {MacDonald}},\
  }\bibfield  {title} {\bibinfo {title} {High-accuracy variational {{Monte
  Carlo}} for frustrated magnets with deep neural networks},\ }\href
  {https://doi.org/10.1103/PhysRevB.108.054410} {\bibfield  {journal} {\bibinfo
   {journal} {Phys. Rev. B}\ }\textbf {\bibinfo {volume} {108}},\ \bibinfo
  {pages} {054410} (\bibinfo {year} {2023})}\BibitemShut {NoStop}%
\bibitem [{\citenamefont {Bernu}\ \emph {et~al.}(1994)\citenamefont {Bernu},
  \citenamefont {Lecheminant}, \citenamefont {Lhuillier},\ and\ \citenamefont
  {Pierre}}]{Bernu1994Triangular}%
  \BibitemOpen
  \bibfield  {author} {\bibinfo {author} {\bibfnamefont {B.}~\bibnamefont
  {Bernu}}, \bibinfo {author} {\bibfnamefont {P.}~\bibnamefont {Lecheminant}},
  \bibinfo {author} {\bibfnamefont {C.}~\bibnamefont {Lhuillier}},\ and\
  \bibinfo {author} {\bibfnamefont {L.}~\bibnamefont {Pierre}},\ }\bibfield
  {title} {\bibinfo {title} {Exact spectra, spin susceptibilities, and order
  parameter of the quantum heisenberg antiferromagnet on the triangular
  lattice},\ }\href {https://doi.org/10.1103/PhysRevB.50.10048} {\bibfield
  {journal} {\bibinfo  {journal} {Phys. Rev. B}\ }\textbf {\bibinfo {volume}
  {50}},\ \bibinfo {pages} {10048} (\bibinfo {year} {1994})}\BibitemShut
  {NoStop}%
\bibitem [{\citenamefont {Reger}\ \emph {et~al.}(1989)\citenamefont {Reger},
  \citenamefont {Riera},\ and\ \citenamefont {Young}}]{Reger89}%
  \BibitemOpen
  \bibfield  {author} {\bibinfo {author} {\bibfnamefont {J.~D.}\ \bibnamefont
  {Reger}}, \bibinfo {author} {\bibfnamefont {J.~A.}\ \bibnamefont {Riera}},\
  and\ \bibinfo {author} {\bibfnamefont {A.~P.}\ \bibnamefont {Young}},\
  }\bibfield  {title} {\bibinfo {title} {{Monte Carlo} simulations of the
  spin-1/2 heisenberg antiferromagnet in two dimensions},\ }\href
  {https://doi.org/10.1088/0953-8984/1/10/007} {\bibfield  {journal} {\bibinfo
  {journal} {J. Phys. Condens. Matter}\ }\textbf {\bibinfo {volume} {1}},\
  \bibinfo {pages} {1855} (\bibinfo {year} {1989})}\BibitemShut {NoStop}%
\bibitem [{\citenamefont {Castro}\ \emph {et~al.}(2006)\citenamefont {Castro},
  \citenamefont {Peres}, \citenamefont {Beach},\ and\ \citenamefont
  {Sandvik}}]{Castro2006HoneycombQMC}%
  \BibitemOpen
  \bibfield  {author} {\bibinfo {author} {\bibfnamefont {E.~V.}\ \bibnamefont
  {Castro}}, \bibinfo {author} {\bibfnamefont {N.~M.~R.}\ \bibnamefont
  {Peres}}, \bibinfo {author} {\bibfnamefont {K.~S.~D.}\ \bibnamefont
  {Beach}},\ and\ \bibinfo {author} {\bibfnamefont {A.~W.}\ \bibnamefont
  {Sandvik}},\ }\bibfield  {title} {\bibinfo {title} {Site dilution of quantum
  spins in the honeycomb lattice},\ }\href
  {https://doi.org/10.1103/PhysRevB.73.054422} {\bibfield  {journal} {\bibinfo
  {journal} {Phys. Rev. B}\ }\textbf {\bibinfo {volume} {73}},\ \bibinfo
  {pages} {054422} (\bibinfo {year} {2006})}\BibitemShut {NoStop}%
\bibitem [{Note1()}]{Note1}%
  \BibitemOpen
  \bibinfo {note} {In a lattice geometry, $C_{ij}$ is translation-invariant, so
  these eigenvectors are plane waves, recovering the usual momentum-space
  treatment.}\BibitemShut {Stop}%
\bibitem [{Note42()}]{Note42}%
  \BibitemOpen
  \bibinfo {note} {Unlike an infinite lattice model, the Goldstone-mode
  operators~\protect \eqref {eq: magnon operator} have a finite support on each
  site, so they do not perfectly commute. However, the norm of the commutator
  is inversely proportional to the system size, so we do not expect it to give
  rise to a low-lying ``antisymmetric two-Goldstone'' excitation. That is, we
  can treat the Goldstone modes as commuting, bosonic operators.}\BibitemShut
  {Stop}%
\bibitem [{\citenamefont {Luttinger}\ and\ \citenamefont
  {Tisza}(1946)}]{LuttingerTisza46}%
  \BibitemOpen
  \bibfield  {author} {\bibinfo {author} {\bibfnamefont {J.~M.}\ \bibnamefont
  {Luttinger}}\ and\ \bibinfo {author} {\bibfnamefont {L.}~\bibnamefont
  {Tisza}},\ }\bibfield  {title} {\bibinfo {title} {Theory of dipole
  interaction in crystals},\ }\href {https://doi.org/10.1103/PhysRev.70.954}
  {\bibfield  {journal} {\bibinfo  {journal} {Phys. Rev.}\ }\textbf {\bibinfo
  {volume} {70}},\ \bibinfo {pages} {954} (\bibinfo {year} {1946})}\BibitemShut
  {NoStop}%
\bibitem [{\citenamefont {Lyons}\ and\ \citenamefont {Kaplan}(1960)}]{Lyons60}%
  \BibitemOpen
  \bibfield  {author} {\bibinfo {author} {\bibfnamefont {D.~H.}\ \bibnamefont
  {Lyons}}\ and\ \bibinfo {author} {\bibfnamefont {T.~A.}\ \bibnamefont
  {Kaplan}},\ }\bibfield  {title} {\bibinfo {title} {Method for determining
  ground-state spin configurations},\ }\href
  {https://doi.org/10.1103/PhysRev.120.1580} {\bibfield  {journal} {\bibinfo
  {journal} {Phys. Rev.}\ }\textbf {\bibinfo {volume} {120}},\ \bibinfo {pages}
  {1580} (\bibinfo {year} {1960})}\BibitemShut {NoStop}%
\bibitem [{\citenamefont {Kaplan}\ and\ \citenamefont
  {Menyuk}(2007)}]{Kaplan07}%
  \BibitemOpen
  \bibfield  {author} {\bibinfo {author} {\bibfnamefont {T.~A.}\ \bibnamefont
  {Kaplan}}\ and\ \bibinfo {author} {\bibfnamefont {N.}~\bibnamefont
  {Menyuk}},\ }\bibfield  {title} {\bibinfo {title} {Spin ordering in
  three-dimensional crystals with strong competing exchange interactions},\
  }\href {https://doi.org/10.1080/14786430601080229} {\bibfield  {journal}
  {\bibinfo  {journal} {Philos. Mag.}\ }\textbf {\bibinfo {volume} {87}},\
  \bibinfo {pages} {3711} (\bibinfo {year} {2007})}\BibitemShut {NoStop}%
\bibitem [{\citenamefont {Cohen}\ and\ \citenamefont
  {Welling}(2016)}]{Cohen2016GroupNetworks}%
  \BibitemOpen
  \bibfield  {author} {\bibinfo {author} {\bibfnamefont {T.~S.}\ \bibnamefont
  {Cohen}}\ and\ \bibinfo {author} {\bibfnamefont {M.}~\bibnamefont
  {Welling}},\ }\bibfield  {title} {\bibinfo {title} {{Group Equivariant
  Convolutional Networks}},\ }\href {https://arxiv.org/abs/1602.07576v3}
  {\bibfield  {journal} {\bibinfo  {journal} {arXiv:1602.07576}\ } (\bibinfo
  {year} {2016})}\BibitemShut {NoStop}%
\bibitem [{\citenamefont {Carleo}\ and\ \citenamefont
  {Troyer}(2017)}]{Carleo2017SolvingNetworks}%
  \BibitemOpen
  \bibfield  {author} {\bibinfo {author} {\bibfnamefont {G.}~\bibnamefont
  {Carleo}}\ and\ \bibinfo {author} {\bibfnamefont {M.}~\bibnamefont
  {Troyer}},\ }\bibfield  {title} {\bibinfo {title} {{Solving the quantum
  many-body problem with artificial neural networks}},\ }\href
  {https://doi.org/10.1126/science.aag2302} {\bibfield  {journal} {\bibinfo
  {journal} {Science}\ }\textbf {\bibinfo {volume} {355}},\ \bibinfo {pages}
  {602} (\bibinfo {year} {2017})}\BibitemShut {NoStop}%
\bibitem [{\citenamefont {Carrasquilla}\ and\ \citenamefont
  {Torlai}(2021)}]{Carrasquilla2021}%
  \BibitemOpen
  \bibfield  {author} {\bibinfo {author} {\bibfnamefont {J.}~\bibnamefont
  {Carrasquilla}}\ and\ \bibinfo {author} {\bibfnamefont {G.}~\bibnamefont
  {Torlai}},\ }\bibfield  {title} {\bibinfo {title} {How to use neural networks
  to investigate quantum many-body physics},\ }\href
  {https://doi.org/10.1103/PRXQuantum.2.040201} {\bibfield  {journal} {\bibinfo
   {journal} {PRX Quantum}\ }\textbf {\bibinfo {volume} {2}},\ \bibinfo {pages}
  {040201} (\bibinfo {year} {2021})}\BibitemShut {NoStop}%
\bibitem [{\citenamefont {Heine}(1960)}]{Heine1960GroupMechanics}%
  \BibitemOpen
  \bibfield  {author} {\bibinfo {author} {\bibfnamefont {V.}~\bibnamefont
  {Heine}},\ }\href@noop {} {\emph {\bibinfo {title} {{Group Theory in Quantum
  Mechanics}}}},\ \bibinfo {series} {International Series in Natural
  Philosophy}, Vol.~\bibinfo {volume} {91}\ (\bibinfo  {publisher} {Pergamon},\
  \bibinfo {address} {Oxford},\ \bibinfo {year} {1960})\BibitemShut {NoStop}%
\bibitem [{\citenamefont {Vicentini}\ \emph {et~al.}(2022)\citenamefont
  {Vicentini}, \citenamefont {Hofmann}, \citenamefont {Szab{\'{o}}},
  \citenamefont {Wu}, \citenamefont {Roth}, \citenamefont {Giuliani},
  \citenamefont {Pescia}, \citenamefont {Nys}, \citenamefont
  {Vargas-Calder{\'{o}}n}, \citenamefont {Astrakhantsev},\ and\ \citenamefont
  {Carleo}}]{Vicentini2022NetKetSystems}%
  \BibitemOpen
  \bibfield  {author} {\bibinfo {author} {\bibfnamefont {F.}~\bibnamefont
  {Vicentini}}, \bibinfo {author} {\bibfnamefont {D.}~\bibnamefont {Hofmann}},
  \bibinfo {author} {\bibfnamefont {A.}~\bibnamefont {Szab{\'{o}}}}, \bibinfo
  {author} {\bibfnamefont {D.}~\bibnamefont {Wu}}, \bibinfo {author}
  {\bibfnamefont {C.}~\bibnamefont {Roth}}, \bibinfo {author} {\bibfnamefont
  {C.}~\bibnamefont {Giuliani}}, \bibinfo {author} {\bibfnamefont
  {G.}~\bibnamefont {Pescia}}, \bibinfo {author} {\bibfnamefont
  {J.}~\bibnamefont {Nys}}, \bibinfo {author} {\bibfnamefont {V.}~\bibnamefont
  {Vargas-Calder{\'{o}}n}}, \bibinfo {author} {\bibfnamefont {N.}~\bibnamefont
  {Astrakhantsev}},\ and\ \bibinfo {author} {\bibfnamefont {G.}~\bibnamefont
  {Carleo}},\ }\bibfield  {title} {\bibinfo {title} {{NetKet 3: Machine
  Learning Toolbox for Many-Body Quantum Systems}},\ }\bibfield  {journal}
  {\bibinfo  {journal} {SciPost Phys. Codebases}\ }\textbf {\bibinfo {volume}
  {7}},\ \href {https://doi.org/10.21468/SCIPOSTPHYSCODEB.7/PDF}
  {10.21468/SCIPOSTPHYSCODEB.7/PDF} (\bibinfo {year} {2022})\BibitemShut
  {NoStop}%
\bibitem [{\citenamefont {Reh}\ \emph {et~al.}(2023)\citenamefont {Reh},
  \citenamefont {Schmitt},\ and\ \citenamefont {Gärttner}}]{Reh2023NQSdesign}%
  \BibitemOpen
  \bibfield  {author} {\bibinfo {author} {\bibfnamefont {M.}~\bibnamefont
  {Reh}}, \bibinfo {author} {\bibfnamefont {M.}~\bibnamefont {Schmitt}},\ and\
  \bibinfo {author} {\bibfnamefont {M.}~\bibnamefont {Gärttner}},\ }\bibfield
  {title} {\bibinfo {title} {Optimizing design choices for neural quantum
  states},\ }\href {https://arxiv.org/abs/2301.06788} {\bibfield  {journal}
  {\bibinfo  {journal} {arXiv:2301.06788}\ } (\bibinfo {year}
  {2023})}\BibitemShut {NoStop}%
\bibitem [{\citenamefont {Klambauer}\ \emph {et~al.}(2017)\citenamefont
  {Klambauer}, \citenamefont {Unterthiner}, \citenamefont {Mayr},\ and\
  \citenamefont {Hochreiter}}]{selu2017}%
  \BibitemOpen
  \bibfield  {author} {\bibinfo {author} {\bibfnamefont {G.}~\bibnamefont
  {Klambauer}}, \bibinfo {author} {\bibfnamefont {T.}~\bibnamefont
  {Unterthiner}}, \bibinfo {author} {\bibfnamefont {A.}~\bibnamefont {Mayr}},\
  and\ \bibinfo {author} {\bibfnamefont {S.}~\bibnamefont {Hochreiter}},\
  }\bibfield  {title} {\bibinfo {title} {Self-normalizing neural networks},\
  }in\ \href
  {https://proceedings.neurips.cc/paper/2017/file/5d44ee6f2c3f71b73125876103c8f6c4-Paper.pdf}
  {\emph {\bibinfo {booktitle} {Advances in Neural Information Processing
  Systems}}},\ Vol.~\bibinfo {volume} {30},\ \bibinfo {editor} {edited by\
  \bibinfo {editor} {\bibfnamefont {I.}~\bibnamefont {Guyon}}, \bibinfo
  {editor} {\bibfnamefont {U.~V.}\ \bibnamefont {Luxburg}}, \bibinfo {editor}
  {\bibfnamefont {S.}~\bibnamefont {Bengio}}, \bibinfo {editor} {\bibfnamefont
  {H.}~\bibnamefont {Wallach}}, \bibinfo {editor} {\bibfnamefont
  {R.}~\bibnamefont {Fergus}}, \bibinfo {editor} {\bibfnamefont
  {S.}~\bibnamefont {Vishwanathan}},\ and\ \bibinfo {editor} {\bibfnamefont
  {R.}~\bibnamefont {Garnett}}}\ (\bibinfo  {publisher} {Curran Associates,
  Inc.},\ \bibinfo {year} {2017})\BibitemShut {NoStop}%
\bibitem [{Note2()}]{Note2}%
  \BibitemOpen
  \bibinfo {note} {This coincidence of 11 states at nearby energies may explain
  the large error bars of the corresponding DMRG result, together with the
  smaller bond dimension (compared to the $S^z=0$ states) used in the
  calculation of Ref.~\cite {Rausch2021}}\BibitemShut {NoStop}%
\bibitem [{\citenamefont {Sandvik}\ and\ \citenamefont
  {Kurkij\"arvi}(1991)}]{Sandvik1991}%
  \BibitemOpen
  \bibfield  {author} {\bibinfo {author} {\bibfnamefont {A.~W.}\ \bibnamefont
  {Sandvik}}\ and\ \bibinfo {author} {\bibfnamefont {J.}~\bibnamefont
  {Kurkij\"arvi}},\ }\bibfield  {title} {\bibinfo {title} {{Quantum Monte Carlo
  simulation method for spin systems}},\ }\href
  {https://doi.org/10.1103/PhysRevB.43.5950} {\bibfield  {journal} {\bibinfo
  {journal} {Phys. Rev. B}\ }\textbf {\bibinfo {volume} {43}},\ \bibinfo
  {pages} {5950} (\bibinfo {year} {1991})}\BibitemShut {NoStop}%
\bibitem [{\citenamefont {Alet}\ \emph {et~al.}(2005)\citenamefont {Alet},
  \citenamefont {Wessel},\ and\ \citenamefont {Troyer}}]{AletSSE}%
  \BibitemOpen
  \bibfield  {author} {\bibinfo {author} {\bibfnamefont {F.}~\bibnamefont
  {Alet}}, \bibinfo {author} {\bibfnamefont {S.}~\bibnamefont {Wessel}},\ and\
  \bibinfo {author} {\bibfnamefont {M.}~\bibnamefont {Troyer}},\ }\bibfield
  {title} {\bibinfo {title} {{Generalized directed loop method for quantum
  Monte Carlo simulations}},\ }\href
  {https://doi.org/10.1103/PhysRevE.71.036706} {\bibfield  {journal} {\bibinfo
  {journal} {Phys. Rev. E}\ }\textbf {\bibinfo {volume} {71}},\ \bibinfo
  {pages} {036706} (\bibinfo {year} {2005})}\BibitemShut {NoStop}%
\bibitem [{\citenamefont {Bauer}\ \emph {et~al.}(2011)\citenamefont {Bauer},
  \citenamefont {Carr}, \citenamefont {Evertz}, \citenamefont {Feiguin},
  \citenamefont {Freire}, \citenamefont {Fuchs}, \citenamefont {Gamper},
  \citenamefont {Gukelberger}, \citenamefont {Gull}, \citenamefont {Guertler},
  \citenamefont {Hehn}, \citenamefont {Igarashi}, \citenamefont {Isakov},
  \citenamefont {Koop}, \citenamefont {Ma}, \citenamefont {Mates},
  \citenamefont {Matsuo}, \citenamefont {Parcollet}, \citenamefont
  {Pawłowski}, \citenamefont {Picon}, \citenamefont {Pollet}, \citenamefont
  {Santos}, \citenamefont {Scarola}, \citenamefont {Schollwöck}, \citenamefont
  {Silva}, \citenamefont {Surer}, \citenamefont {Todo}, \citenamefont {Trebst},
  \citenamefont {Troyer}, \citenamefont {Wall}, \citenamefont {Werner},\ and\
  \citenamefont {Wessel}}]{Alps2}%
  \BibitemOpen
  \bibfield  {author} {\bibinfo {author} {\bibfnamefont {B.}~\bibnamefont
  {Bauer}}, \bibinfo {author} {\bibfnamefont {L.~D.}\ \bibnamefont {Carr}},
  \bibinfo {author} {\bibfnamefont {H.~G.}\ \bibnamefont {Evertz}}, \bibinfo
  {author} {\bibfnamefont {A.}~\bibnamefont {Feiguin}}, \bibinfo {author}
  {\bibfnamefont {J.}~\bibnamefont {Freire}}, \bibinfo {author} {\bibfnamefont
  {S.}~\bibnamefont {Fuchs}}, \bibinfo {author} {\bibfnamefont
  {L.}~\bibnamefont {Gamper}}, \bibinfo {author} {\bibfnamefont
  {J.}~\bibnamefont {Gukelberger}}, \bibinfo {author} {\bibfnamefont
  {E.}~\bibnamefont {Gull}}, \bibinfo {author} {\bibfnamefont {S.}~\bibnamefont
  {Guertler}}, \bibinfo {author} {\bibfnamefont {A.}~\bibnamefont {Hehn}},
  \bibinfo {author} {\bibfnamefont {R.}~\bibnamefont {Igarashi}}, \bibinfo
  {author} {\bibfnamefont {S.~V.}\ \bibnamefont {Isakov}}, \bibinfo {author}
  {\bibfnamefont {D.}~\bibnamefont {Koop}}, \bibinfo {author} {\bibfnamefont
  {P.~N.}\ \bibnamefont {Ma}}, \bibinfo {author} {\bibfnamefont
  {P.}~\bibnamefont {Mates}}, \bibinfo {author} {\bibfnamefont
  {H.}~\bibnamefont {Matsuo}}, \bibinfo {author} {\bibfnamefont
  {O.}~\bibnamefont {Parcollet}}, \bibinfo {author} {\bibfnamefont
  {G.}~\bibnamefont {Pawłowski}}, \bibinfo {author} {\bibfnamefont {J.~D.}\
  \bibnamefont {Picon}}, \bibinfo {author} {\bibfnamefont {L.}~\bibnamefont
  {Pollet}}, \bibinfo {author} {\bibfnamefont {E.}~\bibnamefont {Santos}},
  \bibinfo {author} {\bibfnamefont {V.~W.}\ \bibnamefont {Scarola}}, \bibinfo
  {author} {\bibfnamefont {U.}~\bibnamefont {Schollwöck}}, \bibinfo {author}
  {\bibfnamefont {C.}~\bibnamefont {Silva}}, \bibinfo {author} {\bibfnamefont
  {B.}~\bibnamefont {Surer}}, \bibinfo {author} {\bibfnamefont
  {S.}~\bibnamefont {Todo}}, \bibinfo {author} {\bibfnamefont {S.}~\bibnamefont
  {Trebst}}, \bibinfo {author} {\bibfnamefont {M.}~\bibnamefont {Troyer}},
  \bibinfo {author} {\bibfnamefont {M.~L.}\ \bibnamefont {Wall}}, \bibinfo
  {author} {\bibfnamefont {P.}~\bibnamefont {Werner}},\ and\ \bibinfo {author}
  {\bibfnamefont {S.}~\bibnamefont {Wessel}},\ }\bibfield  {title} {\bibinfo
  {title} {{The ALPS project release 2.0: open source software for strongly
  correlated systems}},\ }\href
  {https://doi.org/10.1088/1742-5468/2011/05/P05001} {\bibfield  {journal}
  {\bibinfo  {journal} {J. Stat. Mech.}\ }\textbf {\bibinfo {volume} {2011}},\
  \bibinfo {pages} {P05001} (\bibinfo {year} {2011})}\BibitemShut {NoStop}%
\bibitem [{\citenamefont {Rausch}\ \emph {et~al.}(2022)\citenamefont {Rausch},
  \citenamefont {Peschke}, \citenamefont {Plorin},\ and\ \citenamefont
  {Karrasch}}]{Rausch2022SOD60}%
  \BibitemOpen
  \bibfield  {author} {\bibinfo {author} {\bibfnamefont {R.}~\bibnamefont
  {Rausch}}, \bibinfo {author} {\bibfnamefont {M.}~\bibnamefont {Peschke}},
  \bibinfo {author} {\bibfnamefont {C.}~\bibnamefont {Plorin}},\ and\ \bibinfo
  {author} {\bibfnamefont {C.}~\bibnamefont {Karrasch}},\ }\bibfield  {title}
  {\bibinfo {title} {{Magnetic properties of a capped kagome molecule with 60
  quantum spins}},\ }\href {https://doi.org/10.21468/SciPostPhys.12.5.143}
  {\bibfield  {journal} {\bibinfo  {journal} {SciPost Phys.}\ }\textbf
  {\bibinfo {volume} {12}},\ \bibinfo {pages} {143} (\bibinfo {year}
  {2022})}\BibitemShut {NoStop}%
\bibitem [{Note3()}]{Note3}%
  \BibitemOpen
  \bibinfo {note} {The Goldstone-operator construction also predicts $\protect
  \mathrm {A_u,A_g}$ quintets. These are indeed at a similar energy as the
  $\protect \mathrm {H_u,H_g}$ ones, but are not as clearly separate from the
  rest of the $S=2$ spectrum as in the C$_{60}$ case.}\BibitemShut {Stop}%
\bibitem [{Note4()}]{Note4}%
  \BibitemOpen
  \bibinfo {note} {This can be seen without detailed calculation: the product
  of three inversion-odd operators is inversion-odd, and the only inversion-odd
  one-dimensional irrep of $I_h$ is $\protect \mathrm {A_u}$}\BibitemShut
  {NoStop}%
\bibitem [{\citenamefont {Konstantinidis}(2023)}]{Konstantinidis2023}%
  \BibitemOpen
  \bibfield  {author} {\bibinfo {author} {\bibfnamefont {N.~P.}\ \bibnamefont
  {Konstantinidis}},\ }\bibfield  {title} {\bibinfo {title} {{Competition
  between frustration and spin dimensionality in the classical
  antiferromagnetic $n$-vector model with arbitrary $n$}},\ }\href
  {https://doi.org/10.21468/SciPostPhysCore.6.2.042} {\bibfield  {journal}
  {\bibinfo  {journal} {SciPost Phys. Core}\ }\textbf {\bibinfo {volume} {6}},\
  \bibinfo {pages} {042} (\bibinfo {year} {2023})}\BibitemShut {NoStop}%
\bibitem [{\citenamefont {King}\ and\ \citenamefont
  {Diudea}(2006)}]{King2006LargeFullerenes}%
  \BibitemOpen
  \bibfield  {author} {\bibinfo {author} {\bibfnamefont {R.~B.}\ \bibnamefont
  {King}}\ and\ \bibinfo {author} {\bibfnamefont {M.~V.}\ \bibnamefont
  {Diudea}},\ }\bibfield  {title} {\bibinfo {title} {The chirality of
  icosahedral fullerenes: a comparison of the tripling (leapfrog), quadrupling
  (chamfering), and septupling (capra) transformations},\ }\href
  {https://doi.org/10.1007/s10910-005-9048-7} {\bibfield  {journal} {\bibinfo
  {journal} {J. Math. Chem.}\ }\textbf {\bibinfo {volume} {39}},\ \bibinfo
  {pages} {597} (\bibinfo {year} {2006})}\BibitemShut {NoStop}%
\bibitem [{\citenamefont {Allen}\ \emph {et~al.}(1995)\citenamefont {Allen},
  \citenamefont {David}, \citenamefont {Fox}, \citenamefont {Ibberson},\ and\
  \citenamefont {Rosseinsky}}]{Allen1995_K3C60_Structure}%
  \BibitemOpen
  \bibfield  {author} {\bibinfo {author} {\bibfnamefont {K.~M.}\ \bibnamefont
  {Allen}}, \bibinfo {author} {\bibfnamefont {W.~I.~F.}\ \bibnamefont {David}},
  \bibinfo {author} {\bibfnamefont {J.~M.}\ \bibnamefont {Fox}}, \bibinfo
  {author} {\bibfnamefont {R.~M.}\ \bibnamefont {Ibberson}},\ and\ \bibinfo
  {author} {\bibfnamefont {M.~J.}\ \bibnamefont {Rosseinsky}},\ }\bibfield
  {title} {\bibinfo {title} {Molecular structure of the fulleride anions in
  superconducting {{K$_3$C$_{60}$}} and insulating {{K$_6$C$_{60}$}} determined
  by powder neutron diffraction},\ }\href {https://doi.org/10.1021/cm00052a023}
  {\bibfield  {journal} {\bibinfo  {journal} {Chem. Mater.}\ }\textbf {\bibinfo
  {volume} {7}},\ \bibinfo {pages} {764} (\bibinfo {year} {1995})}\BibitemShut
  {NoStop}%
\bibitem [{Note5()}]{Note5}%
  \BibitemOpen
  \bibinfo {note} {An example is the C$_{70}$ rugbyballene molecule, whose
  Luttinger--Tisza ground state (in the $\protect \mathrm {A_2'}$ irrep of the
  $D_{5h}$ point group) vanishes on 20 sites near the ends of the molecule,
  requiring an additional $\protect \mathrm {E_2'}$ component to form a valid
  classical ground state. However, this molecule is not inversion symmetric, so
  no chiral state in the strict sense may form on it.}\BibitemShut {Stop}%
\bibitem [{\citenamefont {Bernath}(2005)}]{bernath2005spectra}%
  \BibitemOpen
  \bibfield  {author} {\bibinfo {author} {\bibfnamefont {P.}~\bibnamefont
  {Bernath}},\ }\href@noop {} {\emph {\bibinfo {title} {Spectra of Atoms and
  Molecules}}}\ (\bibinfo  {publisher} {Oxford University Press},\ \bibinfo
  {year} {2005})\BibitemShut {NoStop}%
\bibitem [{\citenamefont {Yamaguchi}\ and\ \citenamefont
  {Yamakage}(2021)}]{Yamaguchi2021MagneticTextureAHE}%
  \BibitemOpen
  \bibfield  {author} {\bibinfo {author} {\bibfnamefont {T.}~\bibnamefont
  {Yamaguchi}}\ and\ \bibinfo {author} {\bibfnamefont {A.}~\bibnamefont
  {Yamakage}},\ }\bibfield  {title} {\bibinfo {title} {Theory of
  magnetic-texture-induced anomalous hall effect on the surface of topological
  insulators},\ }\href {https://doi.org/10.7566/JPSJ.90.063703} {\bibfield
  {journal} {\bibinfo  {journal} {J. Phys. Soc. Jpn.}\ }\textbf {\bibinfo
  {volume} {90}},\ \bibinfo {pages} {063703} (\bibinfo {year}
  {2021})}\BibitemShut {NoStop}%
\bibitem [{\citenamefont {Neuscamman}\ and\ \citenamefont
  {Chan}(2012)}]{Neuscamman2012}%
  \BibitemOpen
  \bibfield  {author} {\bibinfo {author} {\bibfnamefont {E.}~\bibnamefont
  {Neuscamman}}\ and\ \bibinfo {author} {\bibfnamefont {G.~K.-L.}\ \bibnamefont
  {Chan}},\ }\bibfield  {title} {\bibinfo {title} {{Correlator product state
  study of molecular magnetism in the giant Keplerate
  Mo${}_{72}$Fe${}_{30}$}},\ }\href
  {https://doi.org/10.1103/PhysRevB.86.064402} {\bibfield  {journal} {\bibinfo
  {journal} {Phys. Rev. B}\ }\textbf {\bibinfo {volume} {86}},\ \bibinfo
  {pages} {064402} (\bibinfo {year} {2012})}\BibitemShut {NoStop}%
\bibitem [{\citenamefont {Ummethum}\ \emph {et~al.}(2013)\citenamefont
  {Ummethum}, \citenamefont {Schnack},\ and\ \citenamefont
  {L\"auchli}}]{Ummethum2013}%
  \BibitemOpen
  \bibfield  {author} {\bibinfo {author} {\bibfnamefont {J.}~\bibnamefont
  {Ummethum}}, \bibinfo {author} {\bibfnamefont {J.}~\bibnamefont {Schnack}},\
  and\ \bibinfo {author} {\bibfnamefont {A.~M.}\ \bibnamefont {L\"auchli}},\
  }\bibfield  {title} {\bibinfo {title} {{Large-scale numerical investigations
  of the antiferromagnetic Heisenberg icosidodecahedron}},\ }\href
  {https://doi.org/https://doi.org/10.1016/j.jmmm.2012.09.037} {\bibfield
  {journal} {\bibinfo  {journal} {J. Magn. Magn. Mater.}\ }\textbf {\bibinfo
  {volume} {327}},\ \bibinfo {pages} {103} (\bibinfo {year}
  {2013})}\BibitemShut {NoStop}%
\bibitem [{\citenamefont {Cirac}\ \emph {et~al.}(2017)\citenamefont {Cirac},
  \citenamefont {Pérez-García}, \citenamefont {Schuch},\ and\ \citenamefont
  {Verstraete}}]{Cirac2017FundamentalMPS}%
  \BibitemOpen
  \bibfield  {author} {\bibinfo {author} {\bibfnamefont {J.}~\bibnamefont
  {Cirac}}, \bibinfo {author} {\bibfnamefont {D.}~\bibnamefont
  {Pérez-García}}, \bibinfo {author} {\bibfnamefont {N.}~\bibnamefont
  {Schuch}},\ and\ \bibinfo {author} {\bibfnamefont {F.}~\bibnamefont
  {Verstraete}},\ }\bibfield  {title} {\bibinfo {title} {Matrix product density
  operators: Renormalization fixed points and boundary theories},\ }\href
  {https://doi.org/https://doi.org/10.1016/j.aop.2016.12.030} {\bibfield
  {journal} {\bibinfo  {journal} {Ann. Phys.}\ }\textbf {\bibinfo {volume}
  {378}},\ \bibinfo {pages} {100} (\bibinfo {year} {2017})}\BibitemShut
  {NoStop}%
\bibitem [{\citenamefont {Kovesi}(2015)}]{Kovesi2015GoodThem}%
  \BibitemOpen
  \bibfield  {author} {\bibinfo {author} {\bibfnamefont {P.}~\bibnamefont
  {Kovesi}},\ }\bibfield  {title} {\bibinfo {title} {{Good Colour Maps: How to
  Design Them}},\ }\href {http://arxiv.org/abs/1509.03700} {\bibfield
  {journal} {\bibinfo  {journal} {arXiv:1509.03700}\ } (\bibinfo {year}
  {2015})}\BibitemShut {NoStop}%
\end{thebibliography}%

\clearpage

\begin{table}[!t]
    \centering
    \setlength{\tabcolsep}{0.75em}
    \begin{tabular}{cc}
    \begin{tabular}[t]{cc}\hline\hline
        Energy & $S$ \\\hline\hline
        \multicolumn{2}{c}{Irrep $A_1'$}\\[1pt]\hline
        $-15.7336814$ & $0$ \\
        $-15.4531726$ & $1$ \\
        $-15.3587584$ & $2$ \\
        $-15.3407979$ & $0$ \\
        $-15.2635666$ & $0$ \\
        $-15.1351979$ & $1$ \\
        $-15.1042729$ & $1$ \\
        $-15.0863465$ & $0$ \\
        $-15.0376961$ & $2$ \\\hline\hline
        \multicolumn{2}{c}{Irrep $A_2'$}\\[1pt]\hline
        $-15.7736580$ & $1$ \\
        $-15.4543681$ & $1$ \\
        $-15.1821700$ & $1$ \\
        $-15.1214539$ & $1$ \\
        $-15.0740699$ & $0$ \\
        $-15.0340154$ & $2$ \\
        $-15.0087825$ & $1$ \\\hline\hline
        \multicolumn{2}{c}{Irrep $A_1''$}\\[1pt]\hline
        $-15.9372271$ & $0$ \\
        $-15.4928797$ & $0$ \\
        $-15.4621861$ & $0$ \\
        $-15.3572440$ & $2$ \\
        $-15.3195996$ & $1$ \\
        $-15.2629763$ & $1$ \\
        $-15.0493065$ & $1$ \\
        $-15.0184389$ & $2$ \\
        $-15.0035477$ & $1$ \\\hline\hline
        \multicolumn{2}{c}{Irrep $A_2''$}\\[1pt]\hline
        $-15.5004530$ & $1$ \\
        $-15.3936259$ & $1$ \\
        $-15.3823066$ & $1$ \\
        $-15.1635423$ & $1$ \\
        $-15.0894038$ & $1$ \\
        $-15.0325816$ & $2$ \\
        $-15.0287921$ & $0$ \\
        $-15.0202782$ & $2$ \\\hline\hline
    \end{tabular}
    &
    \begin{tabular}[t]{cc}\hline\hline
        Energy & $S$  \\\hline\hline
        \multicolumn{2}{c}{Irrep $E'$}\\\hline
        $-15.8119171$ & $0$ \\
        $-15.6373036$ & $1$ \\
        $-15.6016680$ & $1$ \\
        $-15.2579452$ & $1$ \\
        $-15.2381470$ & $0$ \\
        $-15.2372144$ & $2$ \\
        $-15.1730374$ & $0$ \\
        $-15.1526199$ & $0$ \\
        $-15.1257156$ & $1$ \\
        $-15.1211041$ & $0$ \\
        $-15.0484146$ & $1$ \\
        $-15.0369711$ & $1$ \\
        $-15.0329827$ & $2$ \\
        $-15.0088721$ & $1$ \\\hline\hline
        \multicolumn{2}{c}{Irrep $E''$}\\\hline
        $-15.5748511$ & $0$ \\
        $-15.5658877$ & $1$ \\
        $-15.5407046$ & $1$ \\
        $-15.4218516$ & $0$ \\
        $-15.3801966$ & $1$ \\
        $-15.2871791$ & $1$ \\
        $-15.1992247$ & $2$ \\
        $-15.1629586$ & $1$ \\
        $-15.1445861$ & $0$ \\
        $-15.0995806$ & $1$ \\
        $-15.0982993$ & $0$ \\
        $-15.0617120$ & $0$ \\
        $-15.0514162$ & $1$ \\
        $-15.0397817$ & $2$ \\
        $-15.0141799$ & $2$ \\\hline\hline
    \end{tabular}
    \end{tabular}
    \caption{Energies, spin quantum numbers, and point-group irreps of every eigenstate of the C$_{32}$ Heisenberg model below energy $-15J$ from exact diagonalisation.}
    \label{tab: C32 ED}
\end{table}

\begin{table}[!t]
    \centering
    \setlength{\tabcolsep}{0.75em}
    \begin{tabular}{clll} \hline\hline
        Irrep & \multicolumn{1}{c}{$P=+1$} & \multicolumn{1}{c}{$P=-1$} & \multicolumn{1}{c}{$S^z=2$} \\\hline
        $\mathrm{A_1'}$	&	$-15.7279(3)$	&	$-15.4481(3)$	&	$-15.3491(3)$	\\
        $\mathrm{A_2'}$	&	$-15.0200(5)$\textsuperscript{a}	&	$\mathbf{-15.7648(4)}$	&	$-15.0226(4)$	\\
        $\mathrm{A_1''}$	&	$\mathbf{-15.9342(3)}$	&	$-15.3053(5)$	&	$\mathbf{-15.3537(3)}$	\\
        $\mathrm{A_2''}$	&	$-15.0250(4)$\textsuperscript{a}	&	$-15.4834(5)$	&	$-15.0264(3)$	\\
        $\mathrm{E'}$	&	$-15.8020(4)$	&	$-15.6198(4)$	&	$-15.2220(5)$	\\
        $\mathrm{E''}$	&	$-15.5691(3)$	&	$-15.5591(3)$	&	$-15.1786(6)$	\\\hline\hline
        \multicolumn{4}{l}{\footnotesize\textsuperscript{a}$P=+1$ simulation that returned an $S=2$ state.}
    \end{tabular}
    \caption{Best GCNN variational energies for the C$_{32}$ geometry. Bold numbers correspond to the lowest energy in each spin sector.  We note that the lowest $S=2$ state found by exact diagonalisation belongs to the $A_1'$ irrep; however, the variational energy difference between the $A_1'$ and $A_1''$ sectors is smaller than the difference of either to the exact value.}
    \label{tab: C32 energy}
\end{table}

\begin{table}[!t]
    \centering
    \setlength{\tabcolsep}{0.75em}
    \begin{tabular}{clll} \hline\hline
        Irrep & \multicolumn{1}{c}{$P=+1$} & \multicolumn{1}{c}{$P=-1$} & \multicolumn{1}{c}{$S^z=2$} \\\hline
        $\mathrm{A_g}$	& $\mathbf{-31.1302(2)}$	& $-29.7548(4)$	& $-30.2517(2)$\\
        $\mathrm{A_u}$	& $-29.6807(6)$\textsuperscript{a}	& $-29.8208(7)$	& $-29.6962(4)$\\
        $\mathrm{T_{1g}}$	& $-29.9447(5)$\textsuperscript{a}	& $-30.3186(3)$	& $-29.9378(6)$\\
        $\mathrm{T_{1u}}$	& $-30.1802(3)$	& $-30.2944(3)$	& $-30.0219(5)$\\
        $\mathrm{T_{2g}}$	& $-29.9906(9)$\textsuperscript{a}	& $\mathbf{-30.7685(4)}$	& $-30.0210(4)$\\
        $\mathrm{T_{2u}}$	& $-30.2487(3)$	& $-30.1286(5)$	& $-30.0230(6)$\\
        $\mathrm{G_g}$	& $-30.3033(5)$	& $-30.0876(7)$	& $-30.0850(4)$\\
        $\mathrm{G_u}$	& $-30.3551(6)$	& $-30.6118(5)$	& $-30.1110(3)$\\
        $\mathrm{H_g}$	& $-30.4189(7)$	& $-30.0871(8)$	& $\mathbf{-30.3251(4)}$\\
        $\mathrm{H_u}$	& $-30.2494(6)$\textsuperscript{a}	& $-30.4232(4)$	& $-30.2619(5)$\\\hline
        & \multicolumn{1}{c}{$S=0$} & \multicolumn{1}{c}{$S=1$} & \multicolumn{1}{c}{$S=2$} \\
        DMRG~\cite{Rausch2021} & $-31.131(7)$ & $-30.775(6)$ & $-30.3(2)$ \\
        & $-30.440(9)$  \\\hline\hline
        \multicolumn{4}{l}{\footnotesize\textsuperscript{a}$P=+1$ simulation that returned an $S=2$ state.}
    \end{tabular}
    \caption{Best GCNN variational energies for the C$_{60}$ geometry, compared to the DMRG variational energies of Ref.~\cite{Rausch2021}. 
    Bold numbers correspond to the lowest energy in each spin sector. }
    \label{tab: C60 energy}
\end{table}

\begin{table}[!t]
    \centering
    \setlength{\tabcolsep}{0.75em}
    \begin{tabular}{clll} \hline\hline
        Irrep & \multicolumn{1}{c}{$P=+1$} & \multicolumn{1}{c}{$P=-1$} & \multicolumn{1}{c}{$S^z=2$} \\\hline
        $\mathrm{A_g}$ & $-41.0041(4)$ & $-40.3017(9)$ & $-40.7204(5)$\\
        $\mathrm{A_u}$ & $\mathbf{-41.0387(3)}$ & $-40.6590(5)$ & $-40.6273(5)$\\
        $\mathrm{T_{1g}}$ & $-40.6007(8)$\textsuperscript{a} & $-40.7405(11)$ & $-40.5850(7)$\textsuperscript{b}\\
        $\mathrm{T_{1u}}$ & $-40.7063(8)$\textsuperscript{a} & $-40.8184(9)$ & $-40.7155(7)$\\
        $\mathrm{T_{2g}}$ & $-40.6382(7)$\textsuperscript{c} & $\mathbf{-40.9314(7)}$ & $-40.5858(8)$\\
        $\mathrm{T_{2u}}$ & $-40.5837(9)$\textsuperscript{a} & $-40.8920(9)$ & $-40.6155(9)$\\
        $\mathrm{G_g}$ & $-40.8578(8)$ & $-40.7622(9)$ & $-40.7197(8)$\\
        $\mathrm{G_u}$ & $-40.7444(8)$ & $-40.8820(8)$ & $-40.7117(8)$\\
        $\mathrm{H_g}$ & $-40.8904(12)$ & $-40.8434(6)$ & $-40.7264(10)$\\
        $\mathrm{H_u}$ & $-40.8350(8)$ & $-40.8011(9)$ & $\mathbf{-40.7634(8)}$\\\hline\hline
        \multicolumn{4}{l}{\footnotesize\textsuperscript{a}$P=+1$ simulation that returned an $S=2$ state.}\\[-1ex]
        \multicolumn{4}{l}{\footnotesize\textsuperscript{b}$S^z=2$ simulation that returned an $S=3$ state.}\\[-1ex]
        \multicolumn{4}{l}{\footnotesize\textsuperscript{c}State not clearly dominated by one $S$-sector.}
    \end{tabular}
    \caption{Best GCNN variational energies for the C$_{80}$ geometry. Bold numbers correspond to the lowest energy in each spin sector. }
    \label{tab: C80 energy}
\end{table}

\end{document}